\documentclass[aps,prd,twocolumn,nofootinbib,superscriptaddress]{revtex4-2}

\pdfoutput=1

\usepackage[utf8]{inputenc} 
\usepackage{amsmath}
\usepackage{amssymb}
\usepackage{xcolor}
\usepackage{graphicx}
\usepackage{dcolumn}
\usepackage{bm}
\usepackage{microtype}
\usepackage{threeparttable}
\usepackage{mathtools}
\usepackage[colorlinks=true]{hyperref}
\usepackage{physics}
\usepackage{comment}

\newcommand{\be}{\begin{eqnarray}}
\newcommand{\ee}{\end{eqnarray}}

\begin{document}

\title{Temporal Entanglement from Holographic Entanglement Entropy}

\author{Michal P. Heller}
\email{michal.p.heller@ugent.be}
\affiliation{Department of Physics and Astronomy, Ghent University, 9000 Ghent, Belgium}
\affiliation{Institute of Theoretical Physics and Mark Kac Center for Complex
Systems Research, Jagiellonian University, 30-348 Cracow, Poland}

\author{Fabio Ori}
\email{fabio.ori@ugent.be}
\affiliation{Department of Physics and Astronomy, Ghent University, 9000 Ghent, Belgium}

\author{Alexandre Serantes}
\email{alexandre.serantesrubianes@ugent.be}
\affiliation{Department of Physics and Astronomy, Ghent University, 9000 Ghent, Belgium}

\begin{abstract}

\noindent Recently, several notions of entanglement in time have emerged as a novel frontier in quantum many-body physics, quantum field theory and gravity. We propose a systematic prescription to characterize temporal entanglement in relativistic quantum field theory in a general state for an arbitrary subregion on a flat, constant-time slice in a flat spacetime. Our prescriptions starts with the standard entanglement entropy of a spatial subregion and amounts to transporting the unchanged subregion to boosted time slices all the way across the light cone when it becomes in general a complex characterization of the corresponding temporal subregion. For holographic quantum field theories, our prescription amounts to an analytic continuation of all codimension-two bulk extremal surfaces satisfying the homology constraint and picking the one with the smallest real value of the area as the leading saddle point. We implement this prescription for holographic conformal field theories in thermal states on both a two-dimensional Lorentzian cylinder and three-dimensional Minkowski space, and show that it leads to results with self-consistent physical properties of temporal entanglement.
\end{abstract}

\maketitle

\section{Introduction}

Entanglement and its entropy have been among the very few key notions shaping the development of theoretical physics in the last two decades. Impressive progress ranges from the establishment of entanglement-inferred tensor network algorithms for ab initio simulation of quantum-many body systems on classical computers~\cite{Orus:2013kga,Cirac:2020obd}, through the characterization of topological orders~\cite{Zeng:2015pxf}, shedding light on thermalization of closed quantum systems~\cite{Mori:2017qhg,Berges:2020fwq}, new understanding of irreversibility of renormalization group flows in quantum field theory~\cite{Casini:2022rlv}, all the way to the geometrization of quantum field theory entanglement within the holographic duality~\cite{Nishioka:2009un,Rangamani:2016dms,Chen:2021lnq} and studying its implications for the black hole information paradox~\cite{Almheiri:2020cfm}. All these paradigm-shifting developments stemmed from the standard notion of entanglement entropy associated with a bipartition of quantum systems into spatial subregions on a constant time slice or appropriate algebraic  quantum field theory formalizations of this notion.

While this entanglement in space is arguably quite well understood by now, the notion of entanglement in time is not. In the first place, such a notion is not apparent from the basics of quantum mechanics. Instead, it originates from the field of tensor networks and attempts within it to lower the complexity of algorithms modeling unitary time evolution by devising clever contraction schemes leading to the emergence of the paradigmatic matrix product state tensor networks along the temporal rather than spatial direction~\cite{Banuls:2009jmn,Hastings:2014qqa}. Such a structure allows to define temporal reduced density matrices of several kinds and obtain their characterization in terms of Renyi entropies~\cite{Banuls:2009jmn,Hastings:2014qqa,Frias-Perez:2022wwa,Lerose:2022sxm,Giudice:2021smd}, or their pseudoentropy generalizations to non-Hermitian matrices~\cite{Carignano:2023xbz}.

In the context of relativistic quantum field theories, as originally proposed in~\cite{Doi:2022iyj,Doi:2023zaf} (see also \cite{Narayan:2022afv}), closed-form expressions for single interval entanglement entropy of conformal field theory (CFT) in two spacetime dimensions allow for an explicit analytic continuation to a temporal domain, leading to a notion of timelike entanglement entropy. For example, in the vacuum state in Minkowski space for a single interval of length $\Delta x$ the entanglement entropy reads
\begin{equation}
\label{eq.EECFT2}
S = \frac{c}{3} \log{\frac{\Delta x}{\delta}}\ ,
\end{equation}
where $c$ is the central charge and $\delta$ is the ultraviolet (UV) cut-off~\cite{Holzhey:1994we,Calabrese:2004eu}. The analytic continuation in question amounts to the replacement $\Delta x \rightarrow i \, \Delta t$, yielding a quantity of the same logarithmic dependence but now on $\Delta t$ and having a constant imaginary offset
\begin{equation}
\label{eq.TEECFT2}
S = \frac{c}{3} \log{\frac{\Delta x}{\delta}} + i \frac{\pi}{6}c\,,
\end{equation}
where the principal branch of the logarithm function was picked. Recently, it was shown in one particular example that a similar analytic continuation agrees with the generalization of the von Neumann entropy computed using temporal matrix product states, connecting the two hitherto independent lines of research on temporal entanglement~\cite{Carignano:2024jxb}. In~\cite{Doi:2022iyj,Doi:2023zaf} the quantity encapsulated by~\eqref{eq.TEECFT2} was referred to as timelike entanglement entropy (and for strongly coupled quantum field theories with large number of microscopic constituents as to holographic timelike entanglement entropy).

While our ability to explicitly compute entanglement entropy in a generic quantum field theory is very limited and closed-form expressions like~\eqref{eq.EECFT2} are extremely scarce, for strongly-coupled quantum field theories with large number of microscopic constituents the entanglement entropy has proven surprisingly simple to obtain. In this case, the entanglement entropy is given in terms of the area~$A$ of the extremal surface attached to the spatial subregion of interest lying on the asymptotic boundary of the higher-dimensional geometry where holographic quantum field theories are defined, see~\cite{Ryu:2006bv,Hubeny:2007xt} for the original proposals,~\cite{Casini:2011kv,Lewkowycz:2013nqa,Dong:2016hjy} for approaches to a derivation and~\cite{Nishioka:2009un,Rangamani:2016dms,Chen:2021lnq} for reviews. More precisely, it is given by the associated Bekenstein-Hawking entropy
\begin{equation}
S = \frac{A}{4 G_{N}},
\end{equation}
where $G_{N}$ is the gravitational constant in holography. Given the simplicity of how holography geometrizes entanglement entropy and the aforementioned scarcity of exact expressions for entanglement entropy in other quantum field theories, it is natural to expect that key progress on our understanding of temporal entanglement in quantum field theory will occur through AdS/CFT.

Since temporal entanglement in quantum field theory can be defined by an analytic continuation, it should come as no surprise that holographically the relevant geometric notion will be an analytic continuation of the extremal surfaces geometrizing entanglement entropy, such that they are anchored on a timelike subregion. In~\cite{Heller:2024whi} we identified that such extremal surfaces will be in general complex, i.e. they perceive the bulk geometry for complex rather than real spacetime coordinates. This connects with earlier holographic studies of complex geodesics, which are one-dimensional extremal surfaces, in the context of approximating boundary correlation functions~\cite{Fidkowski:2003nf,Balasubramanian:2012tu,Ceplak:2024bja}.

The key open problem that we address in the present article originates from the existence of multiple complex extremal surfaces satisfying the same boundary condition. This problem appears, for example, in the paradigmatic example of black hole spacetime corresponding to thermal or thermofield double states in dual quantum field theory in three and more spacetime dimensions~\cite{Heller:2024whi}. In the present paper we identify another important instance where there are multiple complex extremal surface candidates to define holographic timelike entanglement entropy.

In the context of entanglement entropy, it is clear how to proceed when there are multiple extremal surfaces~\cite{Headrick:2010zt,Haehl:2014zoa}: all real extremal surfaces homologous to the subregion give saddle point contributions to the holographic entanglement entropy and the one with the smallest area is the dominant one. The remaining saddles give subleading contributions to the holographic entanglement entropy, exponentially suppressed in the difference of areas with respect to the leading saddle. Presence of multiple saddles in the holographic entanglement entropy is responsible, for example, for entanglement entropy being consistent with the cluster decomposition principle of quantum field theory.

What we propose in the present work is a prescription for computing holographic timelike entanglement entropy even if there are multiple nontrivial complex extremal surface candidates. Our guiding principle is that the quantity we define holographically respects the UV-IR correspondence~\cite{Susskind:1998dq}. In particular, the key self-consistency condition will be for us that for sufficiently small temporal subregions in general excited states the quantity they give rise to reduces to the vacuum state answer.

The essence of our proposal is to define timelike entanglement entropy in terms of entanglement entropy by taking the entangling region from spacelike to timelike, going around the light cone in a way illustrated in Fig.~\ref{fig:rotation} that in particular transforms~\eqref{eq.EECFT2} to~\eqref{eq.TEECFT2}. As we show, in holography due to the properties of entanglement entropy for spatial subregions in the vicinity of the light cone there exists an analytic continuation that respects the UV-IR correspondence and that reproduces the vacuum answer for small subsystems in an excited state. As a result, in holography our prescription for timelike entanglement entropy is a natural generalization and a direct consequence of the prescription for holographic entanglement entropy.
\\\\
\noindent \textbf{Note added:} While this work was being finalized, we became aware of the results of Carlos Nunez and Dibakar Roychowdhury~\cite{Nunez:2025ppd}, who also explore spacelike-to-timelike analytic continuations to define timelike entanglement entropy in holography.  We coordinated the submission to appear on the same day on the arXiv.

\section{The key idea \label{sec:idea}}

\subsection{General quantum field theory}

We will be considering quantum field theories primarily in $d$-dimensional Minkowski spacetime. Later in the paper we will also consider theories on a two-dimensional Lorentzian cylinder, where the spatial direction is a circle.

What we are after is a purely Lorentzian approach to defining timelike entanglement entropy. Extrapolating from Eqns.~\eqref{eq.EECFT2} and~\eqref{eq.TEECFT2} and results of~\cite{Doi:2022iyj,Doi:2023zaf} we propose to \emph{define} temporal entanglement entropy by the analytic continuation of standard entanglement entropy as the spacelike region is morphed into a timelike one. 

\begin{figure}[h!]
    \centering    \includegraphics[width=0.7\linewidth]{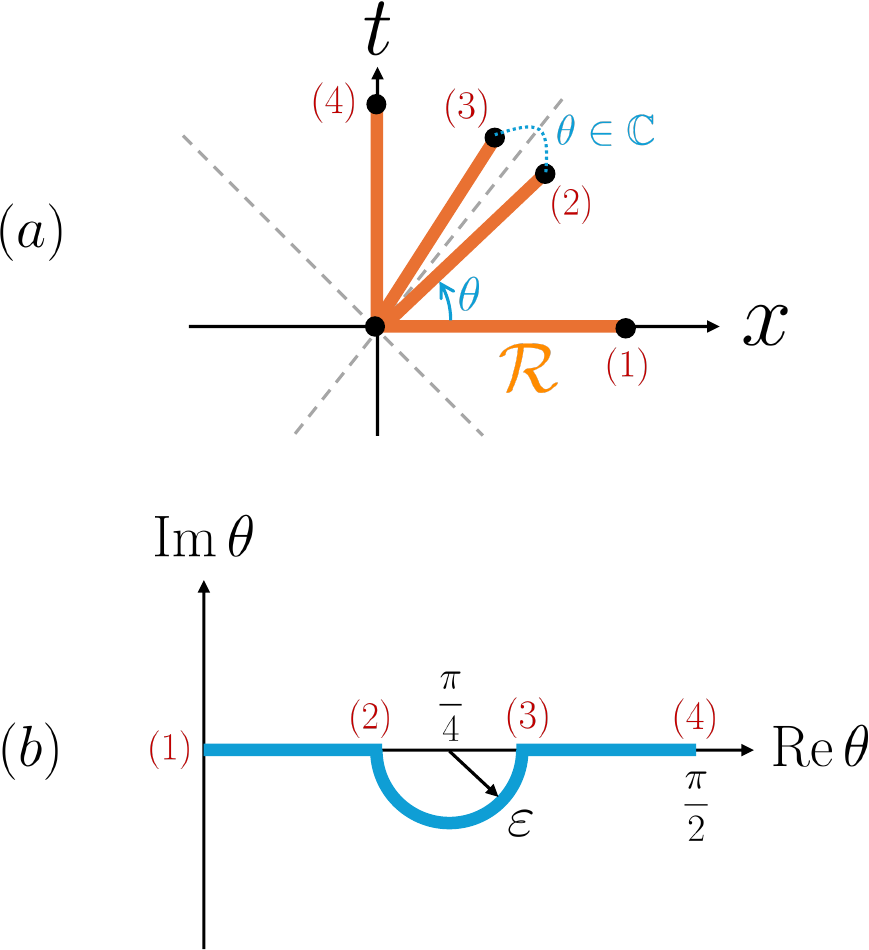}
    \caption{\textbf{(a)} Geometrical analytic continuation of the boundary region $\mathcal{R}$. Starting from a region that lies on a constant time slice (1), the light cone is crossed by slightly complexifying the angle $\theta$ when still in the spacelike regime (2). Once the timelike regime is reached (3), the angle can be increased further to attain purely temporal separations (4). \textbf{(b)}. The trajectory followed by the angle $\theta$ in the analytic continuation. The light cone is crossed by evading the divergence associated with the proper size of the subregion going to~$0$ when $\theta=\frac{\pi}{4}$ with a circle of arbitrarily small radius $\varepsilon$ in the complex $\theta$-plane.}
    \label{fig:rotation}
\end{figure}

Our idea is to keep the embedding of the spacelike subregion on the constant time~$t$ slice fixed and `rotate' this constant time slice in the longitudinal plane spanned by $t$ and a chosen spatial direction $x$, see Fig.~\ref{fig:rotation}(a). The rotation is specified by an angle $\theta$ running between $0$ (the original spatial subregion) and $\frac{\pi}{2}$. Past the light cone located at $\theta = \frac{\pi}{4}$ this produces subregions extending along a timelike direction, as desired.

As in the case of Lorentzian correlation functions obtained from Euclidean correlators, the subtlety lies on the codimension-one hypersurface where the subregion becomes null. There, the proper size of the subregion  
goes to zero and a UV regularization is required. We avoid this singularity in the entanglement entropy by an infinitesimally small detour into complexified Minkowski space. We achieve this by complexifying the rotation angle around $\theta = \frac{\pi}{4}$, see Fig.~\ref{fig:rotation}(b). Coming back to Lorentzian correlators, our prescription can be thought of as a natural generalization of real time $n$-point functions to extended objects: while here we pursue its application to entanglement entropy, the same method could also be applied to, for example, Wilson loops.

The procedure outlined above can be applied to any analytic expression for entanglement entropy, such as~\eqref{eq.EECFT2}, and produce a timelike generalization, such as~\eqref{eq.TEECFT2}. However, given the aforementioned scarcity of such exact results, the true power of our approach lies in allowing to explicitly compute timelike entanglement entropy in holographic setups, which is the focal point of the present article.

\subsection{Holography}
\label{sec:idea_holography}

In holography, when there exists only a single extremal surface satisfying a given asymptotic boundary condition in the spacelike regime---which is, by default, the single contribution to holographic entanglement entropy---the `rotation' outlined in Fig.~\ref{fig:rotation} transforms it into a unique complex extremal surface anchored on a now timelike subregion. This is, for example, the situation in holographic conformal field theories in their vacuum state considered in~\cite{Heller:2024whi}. In this case, the idea of the continuous spacetime transformation of a subregion that we introduced above does not add much new.

The situation changes significantly when multiple extremal surfaces satisfy the homology constraint. As we have previously noted, each of these surfaces should be regarded as a saddle point, and saddle points are known to exchange dominance depending on the parameters that define them. This is well known in the holographic literature and features prominently, for example, in holographic studies of the mutual information~\cite{Headrick:2010zt}. In the case at hand, the varying parameter specifying the saddles (extremal surfaces) is the angle $\theta$, see Fig.~\ref{fig:rotation}. The saddles clearly depend also on the shape of the subregion, but this we decide to keep fixed. 

As a result, the holographic implementation of the spacetime transformation of Fig.~\ref{fig:rotation} in the spacelike region, i.e. for $\theta \leq \frac{\pi}{4} - \varepsilon$, requires to keep track of all the extremal surfaces obeying the homology constraint as a function of $\theta$. In particular, for a fixed subregion shape, this number can change a function of $\theta$. Subsequently all these contributions to the holographic entanglement entropy, the leading and subleading saddles existing at $\theta = \frac{\pi}{4}-\varepsilon$ with $0<\varepsilon \ll 1$, are analytically continued past the light cone to the timelike regime, i.e. $\theta \geq \frac{\pi}{4} + \varepsilon$. In the timelike regime, the dominant contribution to the holographic timelike entanglement entropy comes, as usual, from the saddle that has the smallest real value of the area. Crucially, this does not necessarily imply this will be the contribution that dominates the holographic entanglement entropy. The reason for it is that taking the leading saddle in the spacelike region associated with the limit $G_{N} \rightarrow 0$ does not necessarily commute with the analytic continuation. 

In subsequent sections we will explicitly apply this prescription to the four-dimensional black hole dual to thermal states of three-dimensional holographic conformal field theory in Minkowski space (see Sec.~\ref{sec.R12}) and to the three-dimensional black hole dual to thermal states of two-dimensional conformal field theory on a Lorentzian cylinder (see Sec.~\ref{sec.RS1}).

\subsection{Comments}

It is important to emphasize that the prescription outlined above is valid for \emph{any} state in flat spacetimes and for any flat subregion.

Furthermore, the light cone regulator $\varepsilon$ should be understood in the limiting sense, i.e. $\varepsilon \rightarrow 0^{+}$. Since ultimately in generic cases holographic entanglement entropy is calculated numerically, in practice the limit is probed by taking progressively smaller but nonzero $\varepsilon$ and seeing indications of numerical convergence.

In contrast to the holographic entanglement entropy which one can compute by considering the portion of the bulk limited by future and past pointing light rays emanating from the subregion, our definition of holographic timelike entanglement entropy necessarily requires the understanding of all saddle point contributions to the holographic entanglement entropy right before the light cone is crossed as the parameter $\theta$ is varied and then transforming them to satisfy the desired boundary condition given by the timelike subregion. As a result, holographic timelike entanglement entropy as we define it, at least at this superficial level, requires more information about the bulk than the holographic entanglement entropy.

Moreover, it is easy to understand how the UV-IR correspondence emerges from our prescription. Note that the transformation outlined in Fig.~\ref{fig:rotation} keeps the shape of the subregion intact. As the light cone is approached from the spacelike domain where the quantity one computes is holographic entanglement entropy, the proper size of the subregion along one of the directions goes to zero as a result of the Lorentz contraction. This makes the subregion a very thin slab for which one expects the extremal surface that gives the dominant contribution to the holographic entanglement entropy to lie very close to the boundary. Subsequently this universal contribution sensitive to the vacuum physics is analytically continued to the timelike regime.

Also, let us emphasize that keeping the shape of the subregion rigid during the `rotations', which we view as a natural condition to impose, gets rid of most of the potential ambiguities in the analytic continuation of the entanglement entropy. The only one remaining is related to going `below' or `above' the light cone, see Fig.~\ref{fig:rotation}. This should be contrasted with correlators of local operators where analytic continuation from a constant time slice suffers from substantial (and natural) ambiguities related to the ordering of operators.

Finally, we want to acknowledge that earlier works that study the changes of entanglement entropy under `rotations' include~\cite{Kusuki:2017jxh} and~\cite{Liu:2022ugc}. While~\cite{Kusuki:2017jxh} focused on spacelike slices, the results in~\cite{Liu:2022ugc} involve analogues of spacetime `rotations' in quantum spin chains in one spatial dimension. In particular, the latter indicates our prescription can be systematically studied in quantum many-body systems using tensor network methods or focusing on Gaussian states.

\section{Holographic thermal state on $\mathbb{R}^{1,2}$ \label{sec.R12}} 

In the present section we will employ the prescription advocated in Sec.~\ref{sec:idea} to interpret the multiple complex extremal surfaces from Ref.~\cite{Heller:2024whi}. The paradigmatic setting in question consists of a four-dimensional black brane spacetime and a strip subregion on the boundary, see Fig.~\ref{fig:striprotation}. The holographic entanglement entropy for a strip subregion with $\theta = 0$ has been considered in  Ref.~\cite{Erdmenger:2017pfh}, and  the holographic timelike entanglement entropy candidate extremal surfaces for a strip subregion with $\theta = \frac{\pi}{2}$ in Ref.~\cite{Heller:2024whi}. In the present section we will morph the results from $\theta = 0$ into $\theta = \frac{\pi}{2}$ and show that, in the timelike regime, our prescription picks the extremal surfaces that fulfill the UV--IR correspondence. In particular, it will forbid one class of extremal surfaces from contributing for sufficiently small subregions.

\subsection{Setup}

\begin{figure}[h!]
    \centering    
\includegraphics[width=0.7\linewidth]{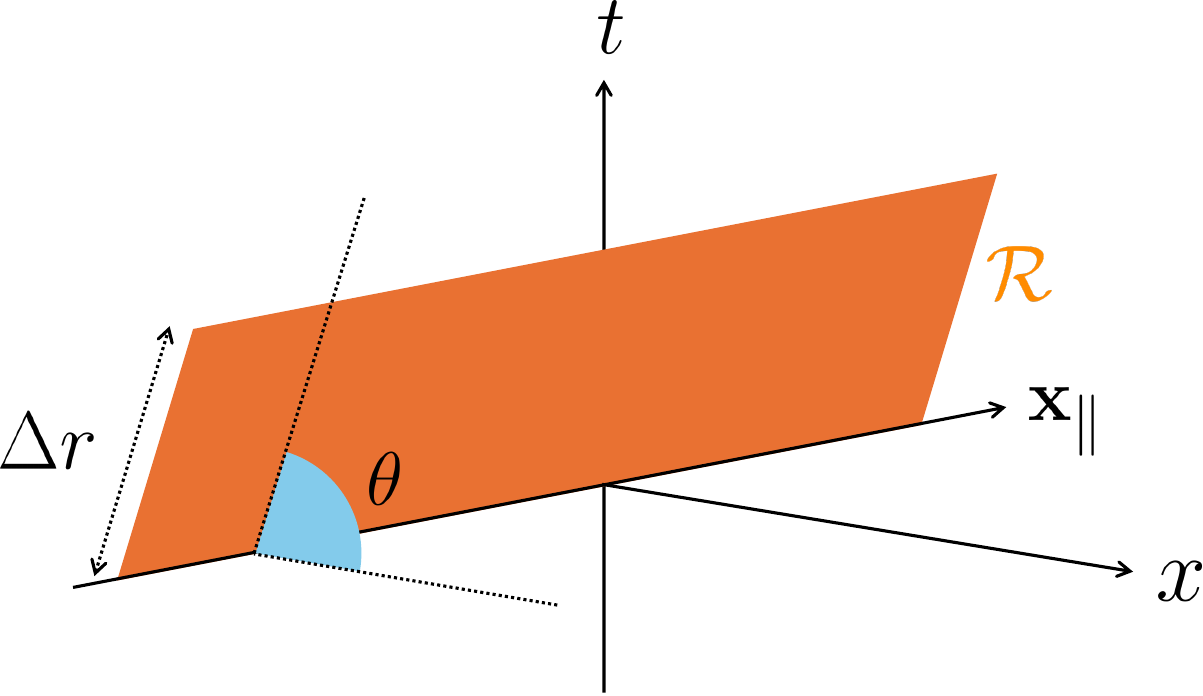}
    \caption{Geometry of a strip boundary subregion $\mathcal{R}$ in three-dimensional Minkowski space. The strip is rotated in the $t$-$x$ plane keeping the coordinate extent $\Delta r$ fixed. See Fig.~\ref{fig:rotation} for the case of a general subregion.}
    \label{fig:striprotation}
\end{figure}

The strip subregion of interest lives in the three-dimensional Minkowski spacetime located at the $z=0$ asymptotic boundary of the four-dimensional bulk geometry 
\begin{equation}\label{metric}
ds^2 = \frac{1}{z^2}\left(\frac{dz^2}{f(z)} - f(z) dt^2  + d\textbf{x}^2 \right),        
\end{equation}
where we have set the curvature scale to one. The choice $f(z) = 1$ corresponds to the empty anti-de Sitter (AdS) space, dual to the vacuum state of the boundary conformal field theory, whereas $f(z) = 1 - \frac{z^3}{z_H^3}$ corresponds to a black brane, dual a thermal state. Splitting the boundary spatial coordinates as $\textbf{x} = \{x, x_\parallel \}$, the strip is defined as 
\begin{equation}\label{strip}
\begin{split}
\mathcal{R} \equiv &\left\{(t,\textbf{x}): t{=}r \sin\theta, x{=}r \cos\theta, r \in \left[0,\Delta r\right], x_\parallel \in \mathbb{R}\right\},
\end{split}
\end{equation}
where $\theta$ is fixed. The projection of the strip $\mathcal{R}$ on the $x$--$t$ plane is a segment joining the origin with the point $(\Delta r \cos\theta, \Delta r\sin\theta)$. The strip is spacelike for $\theta \in [0,\frac{\pi}{4})$, null for $\theta  = \frac{\pi}{4}$, and timelike for $\theta \in (\frac{\pi}{4},\frac{\pi}{2}]$. The case considered in  Ref.~\cite{Heller:2024whi} corresponds to $\theta = \frac{\pi}{2}$. See Fig.~\ref{fig:striprotation} for an illustration of the setup.

By symmetry, the codimension-two bulk extremal surface $\gamma_\mathcal{R}$ takes the form 
\begin{equation}\label{extremal_surface}
X^\mu(\lambda) = \{z_s(\lambda), t_s(\lambda),x_s(\lambda),x_\parallel\},    
\end{equation}
where $\lambda$ is a parameter moving along the variable part of the surface. Given this, we need to extremize the area density functional, 
\begin{equation}
\label{action}
\mathcal{A} \equiv \frac{A}{V} \equiv \int d\lambda \mathcal{L} \equiv \int d\lambda \sqrt{\frac{\frac{z_s'^2}{f(z_s)}{-}f(z_s)t_s'^2{+} x_s'^2}{z_s^4}},   
\end{equation}
to find the entropy density $\mathcal{S} \equiv \mathcal{A}/(4 G)$. In this expression, $V$ stands for the (formally infinite) volume of the line spanned by $x_\parallel$. 

The area density \eqref{action} is a UV-divergent quantity. In the following, we will extract this UV divergence and work with the regularized area density 
\begin{equation}\label{area_density_reg}
\mathcal{A}_\textrm{reg} \equiv \lim_{\delta \to 0 }\left(\mathcal{A} - \frac{2}{\delta}\right),     
\end{equation}
where $\delta \ll 1$ corresponds to the location the regularized asymptotic boundary in the radial direction of the bulk spacetime, $z=\delta$. Correspondingly, we also define 
\begin{equation}
\mathcal{S}_\textrm{reg} \equiv \mathcal{A}_\textrm{reg}/(4 G)    
\end{equation} 
as the regularized entropy density. 

\subsection{Holographic entanglement entropy for a horizontal strip}

\begin{figure}[h!]
\begin{center}
\includegraphics[width=.49\linewidth]{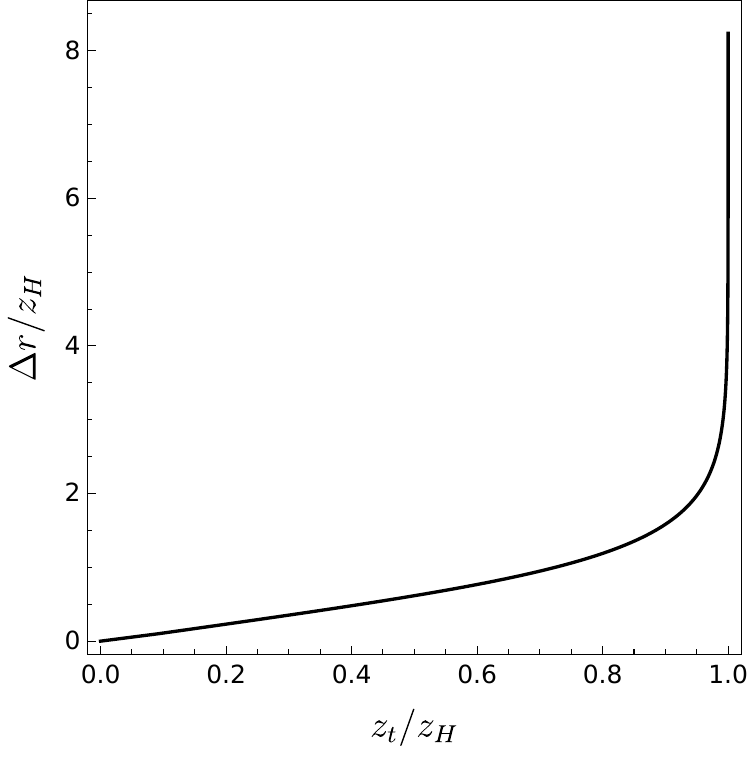}
\includegraphics[width=.49\linewidth]{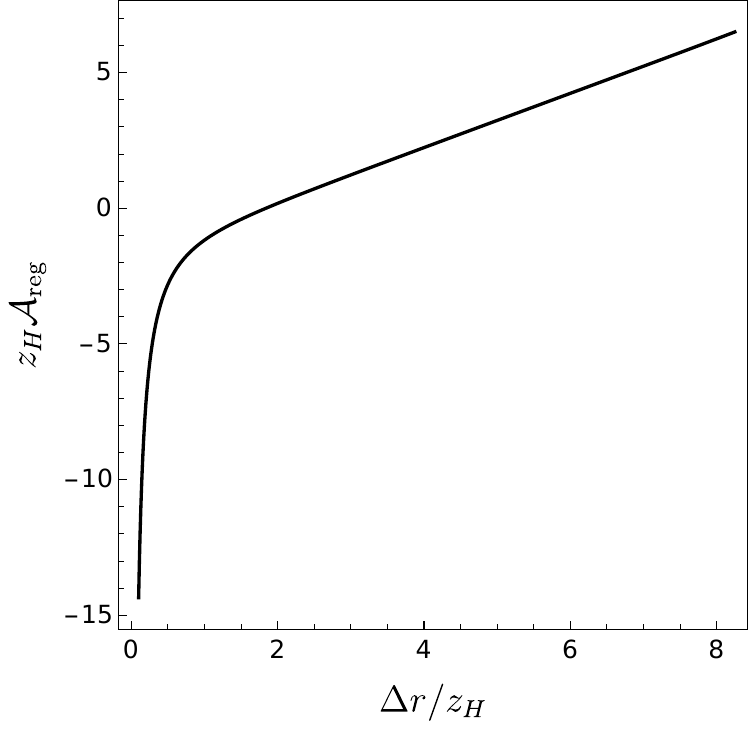}
\caption{\small \textbf{Left panel:} width of the  strip $\Delta r$ as a function of the position of the entangling surface tip $z_t$ in the bulk spacetime. \textbf{Right panel:} regularized area density of the strip $\mathcal{A}_\textrm{reg}$ as a function of the strip width $\Delta r$.} 
\label{fig:all_theta=0}
\end{center}
\end{figure}

The calculation of the entanglement entropy of a spacelike strip with $\theta = 0$ is a standard problem in holography. One finds that, for a given width $\Delta r$, there is a single real entangling surface. In the $\Delta r \to 0$ limit, the tip $z_t$ of this entangling surface approaches the asymptotic boundary $z=0$, while in the opposite $\Delta r \to \infty$ limit the tip approaches the black brane horizon $z=z_H$. Correspondingly, for $\Delta r \to 0$ the regularized entanglement entropy approaches its value in the vacuum state and diverges as $\Delta r^{-1}$, while for $\Delta r \to \infty$ it grows linearly in $\Delta r$ with a slope governed by the location of the event horizon. See Fig.~\ref{fig:all_theta=0} for an illustration of these facts. 

\subsection{Holographic timelike entanglement entropy for a vertical strip}

\begin{figure}[h!]
\begin{center}
\includegraphics[width=\linewidth]{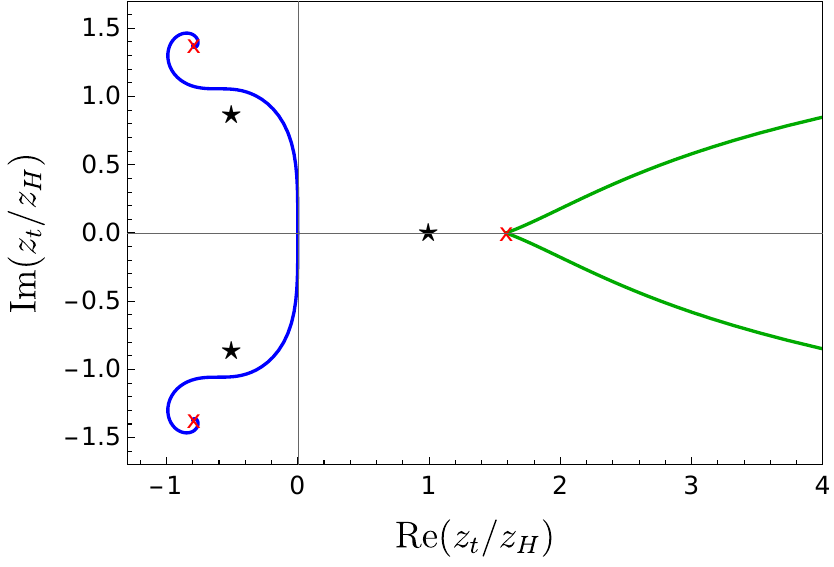}
\caption{\small For a strip with $\theta=\frac{\pi}{2}$, $z_t$ for all the known complex extremal hypersurfaces in an AdS$_4$-Schwarzschild black brane. Blue (green) curves correspond to vacuum-connected (vacuum-disconnected) solutions. Horizons understood as roots of $f(z) = 0$, see Eq.~\eqref{metric}, are represented as black stars, and critical extremal surfaces  as red crosses. This plot appeared earlier in Ref.~\cite{Heller:2024whi}.} 
\label{fig:solution_space_Pi2}
\end{center}
\end{figure}

Ref.~\cite{Heller:2024whi} studied the holographic timelike entanglement entropy for timelike strips with $\theta = \frac{\pi}{2}$. One of the main findings of~\cite{Heller:2024whi} was that the space of complex extremal surfaces associated to this boundary subregion comprises two classes of solutions, referred to as vacuum-connected and vacuum-disconnected. Each class of solutions consists of two branches related to each other by complex conjugation. See Fig.~\ref{fig:solution_space_Pi2} for the location of these branches in the complex $z_t$-plane. 

The main properties of these two classes of solutions were as follows: 
\begin{itemize}
\item For $\Delta r \to 0$, the tips of the vacuum-connected branches flow to the location of the asymptotic boundary $z_t\to0$, while the tips of the vacuum-disconnected ones flow to $|z_t|\to\infty$. Correspondingly, in the $\Delta r \to 0$ limit the regularized area density of the vacuum-connected branches approaches the purely imaginary vacuum answer \cite{Doi:2023zaf}, while the regularized area density of the vacuum-disconnected ones goes to a complex constant with negative real part. 
\item For $\Delta r \to \infty$, the tips of each branch flow to the location of a critical extremal surface for which $z_s(\lambda) = z_c \in \mathbb C$.  This entails that, in the $\Delta r \to \infty$ limit, the regularized area density of each branch grows linearly in $\Delta r$, with a slope determined by the corresponding critical extremal surface.
\end{itemize} 
The key problem left open by Ref.~\cite{Heller:2024whi} was how these different classes of complex extremal surfaces contribute to the holographic timelike entanglement entropy. In particular, note that minimizing over $\Re \mathcal{A}_\textrm{reg}$ to select the dominant saddle would entail that, in the $\Delta r \to 0$ limit, the relevant solutions are the vacuum-disconnected ones and hence the holographic timelike entanglement entropy thus defined does not reduce to the vacuum answer. Armed with the prescription put forward in Sec.~\ref{sec:idea}, we will return to this crucial question in Sec.~\ref{sec:black_brane_TEE} below. 

\subsection{Entanglement entropy in the vicinity of the light cone} 

One of the key questions we have to address is how, under the analytic continuation described in Sec.~\ref{sec:idea}, the single branch of real extremal surfaces associated to a spacelike strip with $\theta = 0$ gives way to the four branches of complex extremal surfaces associated to a timelike strip with $\theta = \frac{\pi}{2}$. To start answering this question, in this subsection we explore the behavior of the holographic entanglement entropy when a spatial strip with $\theta < \frac{\pi}{4}$ approaches the null limit~$\theta = \frac{\pi}{4}$.

Our first main result is that, as $\theta \to \frac{\pi}{4}$, past a critical angle $\theta_c$ there exists a first-order phase transition in the entanglement entropy as $\Delta r$ increases at fixed $\theta$. At this first-order phase transition, the entangling surface $\gamma_\mathcal{R}$ changes discontinuously and its tip $z_t$ goes from being located close to the asymptotic boundary to being located close to the event horizon. 
\begin{figure}[h!]
    \centering    \includegraphics[width=\linewidth]{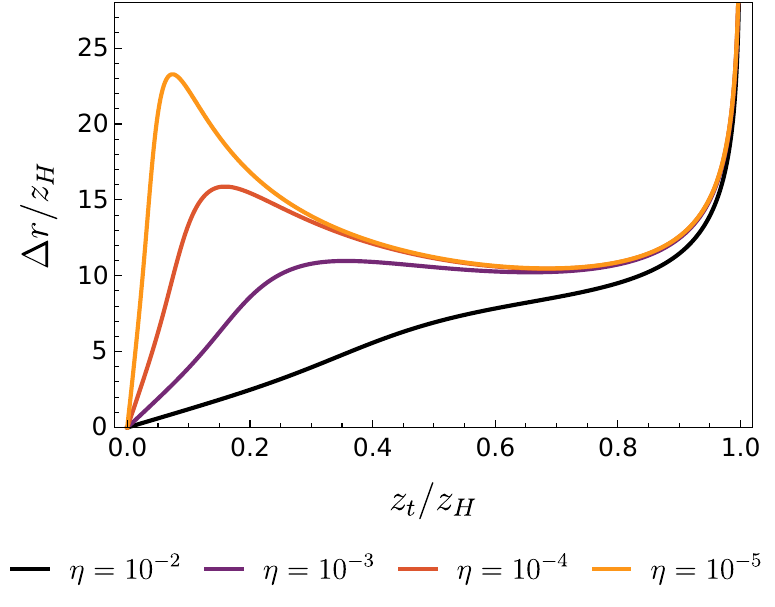}\\\includegraphics[width=\linewidth]{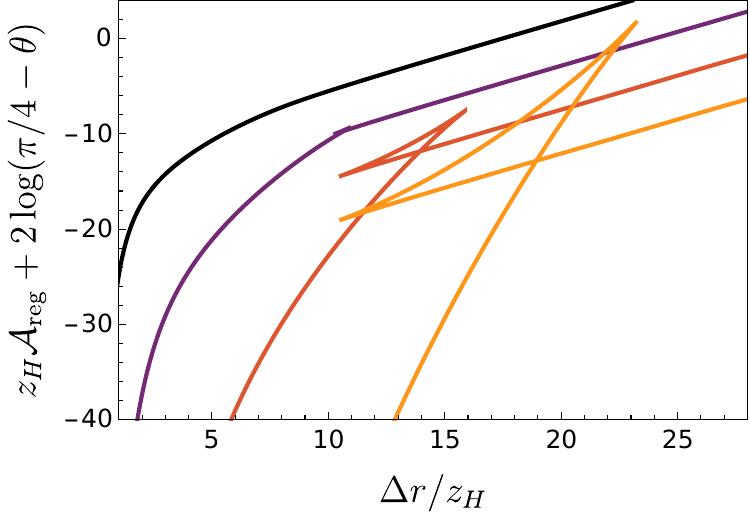} 
    \caption{\textbf{Top panel}: width of the strip $\Delta r$ as a function of the tip of the extremal surface $z_t$ for various values of $\theta = \cot^{-1}(1+\eta)$. \textbf{Bottom panel}: regularized area density $\mathcal{A}_\textrm{reg}$ as a function of $\Delta r$ for various angles. Each $\mathcal{A}_\textrm{reg}$ has been shifted by $\log(\frac{\pi}{4} - \theta)^2$ to prevent the curves from overlapping at large $\Delta r$.}
    \label{fig:phase_transition}
\end{figure}

In the top panel of Fig.~\ref{fig:phase_transition}, we plot $\Delta r$ as a function of $z_t$ for values of $\theta$ progressively closer to $\frac{\pi}{4}$. We clearly see that, as $\varepsilon = \frac{\pi}{4} - \theta \to 0$, $\Delta r$ transitions from being a monotonic function of $z_t$ to having a local maximum and a local minimum. Let the local maximum and minimum be respectively associated with widths $\Delta r_\textrm{max}(\varepsilon)$, $\Delta r_\textrm{min}(\varepsilon)$ and tips $z_{t,\textrm{max}}(\varepsilon)$, $z_{t,\textrm{min}}(\varepsilon)$. These extrema naturally divide the entangling surface candidates into three branches: 
\begin{itemize}
\item \emph{Vacuum-connected}, with $0 \leq z_t < z_{t,\textrm{max}}(\varepsilon)$. 
\item \emph{Unstable}, with $z_{t,\textrm{max}}(\varepsilon) \leq z_t < z_{t,\textrm{min}}(\varepsilon)$.
\item \emph{Horizon-connected}, with $z_{t,\textrm{min}}(\varepsilon) \leq z_t < z_H$.  
\end{itemize}
For strips with $\Delta r < \Delta r_\textrm{min}(\varepsilon)$ or $\Delta r > \Delta r_\textrm{max}(\varepsilon)$ there is a single extremal surface that can contribute to the entanglement entropy, while for strips with $\Delta r \in [\Delta r_\textrm{min}(\varepsilon), \Delta r_\textrm{max}(\varepsilon)]$ there are several. In the latter case, the holographic entanglement entropy prescription instructs us to select the one with the minimal area density, as all of them do obey the homology constraint. As $\Delta r$ increases, this competition leads to a first-order phase transition where the entangling surface jumps from the vacuum-connected branch to the horizon-connected one at a critical separation $\Delta r_c(\varepsilon) \in [\Delta r_\textrm{min}(\varepsilon), \Delta r_\textrm{max}(\varepsilon)]$. The unstable branch is always subdominant. See the bottom panel of Fig.~\ref{fig:phase_transition} for a plot of the regularized area density $\mathcal{A}_\textrm{reg}$ as a function of $\Delta r$ for various angles. 

\begin{figure}[h!]
    \centering    \includegraphics[width=0.95\linewidth]{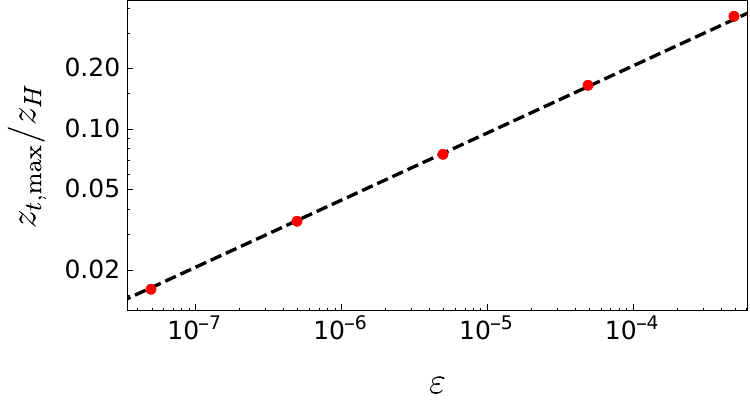}\\\includegraphics[width=0.95\linewidth]{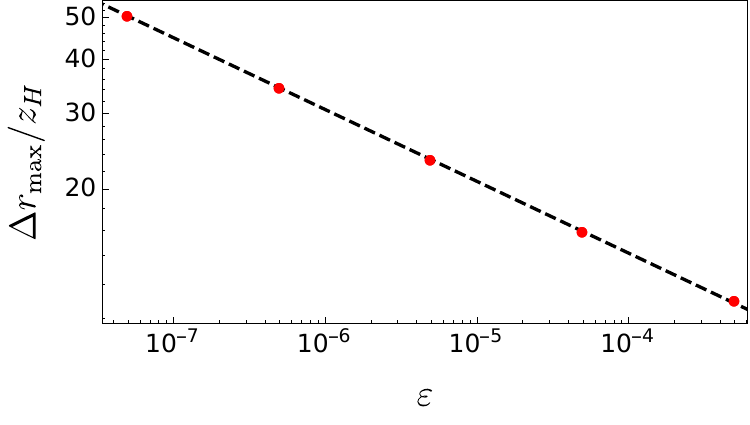} 
    \caption{\textbf{Top panel}: $z_{t,\textrm{max}}/z_H$ as a function of $\varepsilon$ (red dots) together with a fit to a $\varepsilon^{1/3}$ power-law (black dashed line). 
    \textbf{Bottom panel}: $\Delta r_\textrm{max}/z_H$ as a function of $\varepsilon$ (red dots) together with a fit to a $\varepsilon^{-1/6}$ power-law (black dashed line).}
    \label{fig:phase_transition_2}
\end{figure}

Our numerical results are compatible with the fact that, as $\varepsilon \to 0$ and the null limit is approached, $z_{t,\textrm{min}}(\varepsilon)$ and $\Delta r_\textrm{min}(\varepsilon)$ saturate, while $z_{t,\textrm{max}}(\varepsilon)\to0$ and $\Delta r_\textrm{max}(\varepsilon)\to\infty$ (see Fig.~\ref{fig:phase_transition_2}). Hence, the window of widths for which a first-order phase transition is possible, $\Delta r \in [\Delta r_\textrm{min}(\varepsilon), \Delta r_\textrm{max}(\varepsilon)]$, is bounded from below and unbounded from above as $\varepsilon \to 0$. The critical width at which the phase transition itself takes place, $\Delta r_c(\varepsilon)$, also diverges in the same limit. For future reference, we define $\Delta r_\textrm{min}^\star \equiv \lim_{\varepsilon\to0} \Delta r_\textrm{min}(\varepsilon) \approx 10.486 z_H$

Note that, in the light of these results, for a given $\Delta r$ it is always possible to pick $\theta$ sufficiently close to $\frac{\pi}{4}$ such that $\frac{\Delta r}{\Delta r_c}$ is arbitrarily small, $\frac{z_t}{z_H}$ is arbitrarily close to zero and, as a consequence, the regularized entanglement entropy $\mathcal{S}_\textrm{reg}$ is arbitrarily close to its vacuum value. This can be understood as a manifestation of the UV-IR correspondence at the level of the entanglement entropy, since in the $\theta \to \frac{\pi}{4}^-$ limit, the proper width of a strip with fixed $\Delta r$ goes to zero. 

\begin{figure}[h!]
    \centering    
    \includegraphics[width=\linewidth]{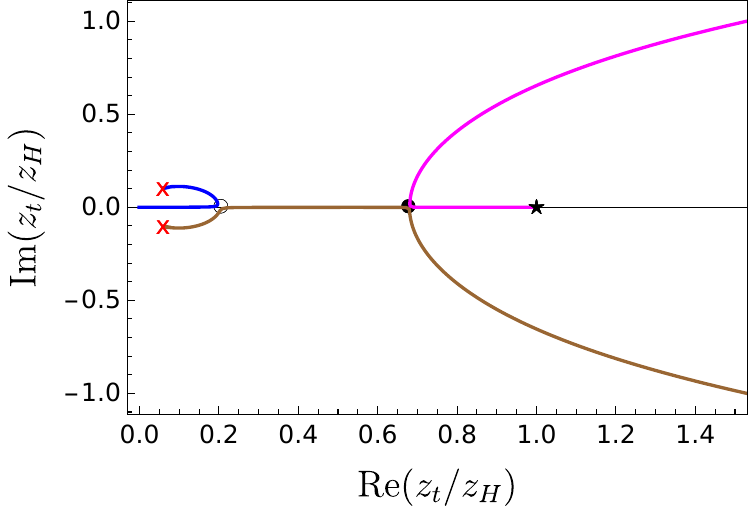}
    \caption{Relevant branches of complex extremal surfaces for $\theta = \frac{\pi}{4}-10^{-4}-10^{-6} i$ as $\Delta r$ is varied. The vacuum-connected, unstable and horizon-connected branches of real extremal surfaces now respectively belong to the blue, brown and magenta branches of complexified solutions. The event horizon is denoted by a black star, the location of $z_{t,\textrm{max}}$ ($z_{t,\textrm{min}}$) for $\theta = \frac{\pi}{4}-10^{-4}$ by open (filled) circles, and the tips of the critical extremal surfaces to which the complexified vacuum-connected and unstable branches flow as $\Delta r \to \infty$ by red crosses.
    }
    \label{fig:solution_space_theta_complex}
\end{figure}

Finally, we would like to point out that the first-order phase transition we have uncovered can be understood as emerging from a collision between real and complex branches of extremal surfaces. This fact follows from the observation that, even though for $\theta < \frac{\pi}{4}$ the boundary subregion is spacelike, there still exist complex-conjugated branches of complex extremal surfaces emanating from both $z_{t,\textrm{max}}$ and $z_{t,\textrm{min}}$. Moving away from $z_{t,\textrm{max}}$ along these complex branches takes one to progressively larger $\Delta r \geq \Delta r_\textrm{max}$, while moving away from $z_{t,\textrm{min}}$ along them takes one to progressively smaller $\Delta r \leq \Delta r_\textrm{min}$. Upon complexification of the angle $\theta$, the branch collisions are resolved and one finally obtains three smooth branches of extremal surfaces, which are now correspondingly complex. For $\Im(\theta)\to0^-$, the branch rearrangement is as follows: 
\begin{itemize}
    \item The vacuum-connected branch of real extremal surfaces merges with the upper branch of complex extremal surfaces emanating from $z_{t,\textrm{max}}$.
    \item The lower branch of complex extremal surfaces emanating from $z_{t,\textrm{max}}$, the unstable branch of real extremal surfaces, and the lower branch of complex extremal surfaces emanating from $z_{t,\textrm{min}}$ all merge.
    \item The horizon-connected branch of real extremal surfaces merges with the upper branch of complex extremal surfaces emanating from $z_{t,\textrm{min}}$.
\end{itemize}
See Fig.~\ref{fig:solution_space_theta_complex} for an example. 

\subsection{Analytical continuation past the light cone}

According to the prescription put forward in Sec.~\ref{sec:idea}, to compute the timelike entanglement entropy of a strip with width $\Delta r$ and tilt $\theta > \frac{\pi}{4}$, we first have to select the real extremal surfaces contributing to the entanglement entropy of a strip with the same width and $\theta = \frac{\pi}{4}-0^+$, then analytically continue these real extremal surfaces around the light cone, and finally pick among the resulting complex extremal surfaces the one with the smallest real part of the area density. 

Our choice of analytical continuation is encapsulated in Fig.~\ref{fig:rotation}(b). We fix $\Delta  r$ and choose $\theta = \frac{\pi}{4} - \varepsilon$ with $\varepsilon = 0^+$. Then, we set 
\begin{equation}\label{alpha_complex}
\theta = \frac{\pi}{4} - \varepsilon \, e^{i\alpha}, 
\end{equation}
keep $\varepsilon$ fixed, and follow the initial real extremal surface as $\alpha$ goes from $0$ to $\pi$. The end result is a complex extremal surface associated to a timelike strip with $\theta = \frac{\pi}{4} + \varepsilon$. Finally, we follow this complex extremal surface as $\theta$ goes from $\frac{\pi}{4} +\varepsilon$ to $\frac{\pi}{2}$. We carry out this procedure numerically by working with a small-but-finite~$\varepsilon$. We note that, since for a given $\Delta r$ is always possible to choose $\varepsilon$ sufficiently small so that $\Delta r < \Delta r_\textrm{max}(\varepsilon)$, below we will only consider the cases where $\Delta r < \Delta r_{min}(\varepsilon)$ and $\Delta r \in [\Delta r_\textrm{min}(\varepsilon),\Delta r_\textrm{max}(\varepsilon)]$. 
\\\\
\underline{$\Delta r < \Delta r_{min}(\varepsilon)$}.~In this case, only real extremal surfaces belonging to the vacuum-connected branch can contribute to the holographic entanglement entropy for $\theta\to\frac{\pi}{4}^-$. Our second main result is that, for this branch, the analytic continuation described above maps the initial real extremal surface at $\theta<\frac{\pi}{4}$ to a solution at $\theta = \frac{\pi}{2}$ that belongs to the upper vacuum-connected branch of complex extremal surfaces, depicted in blue in Fig.~\ref{fig:solution_space_Pi2}. Crucially, this implies that, for $\Delta r < \Delta r_{min}(\varepsilon)$, the vacuum-disconnected branches of complex extremal surfaces at $\theta=\frac{\pi}{2}$, depicted in green in Fig.~\ref{fig:solution_space_Pi2}, do not correspond to the analytic continuation of real entangling surfaces and hence cannot contribute to the timelike entanglement entropy as we have defined it. This fact is critical for our holographic timelike entanglement entropy prescription to uphold the UV--IR correspondence. 
\begin{figure}[h!]
     \centering     \includegraphics[width=\linewidth]{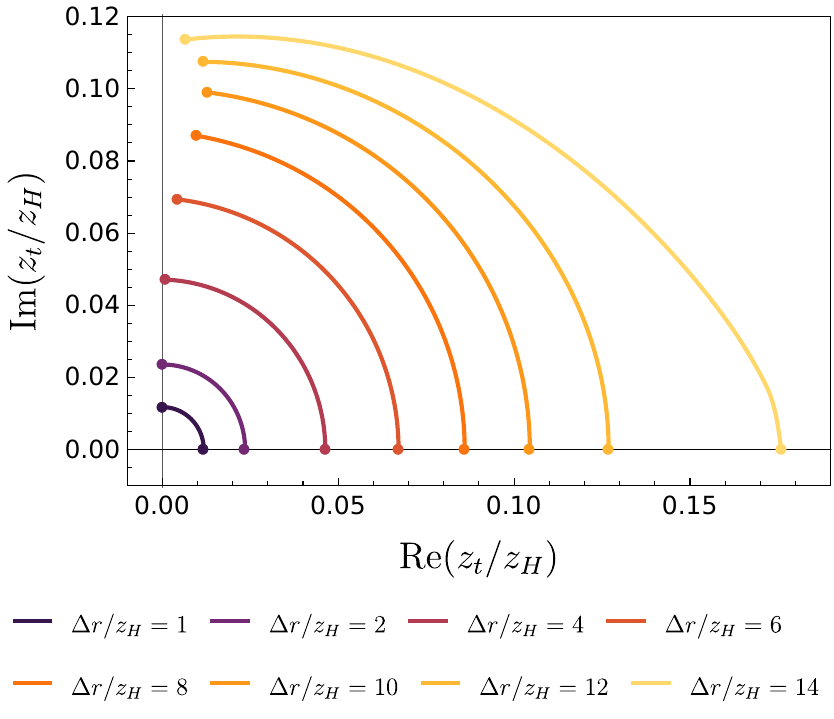}\\
     \vspace{0em}
     \includegraphics[width=\linewidth]{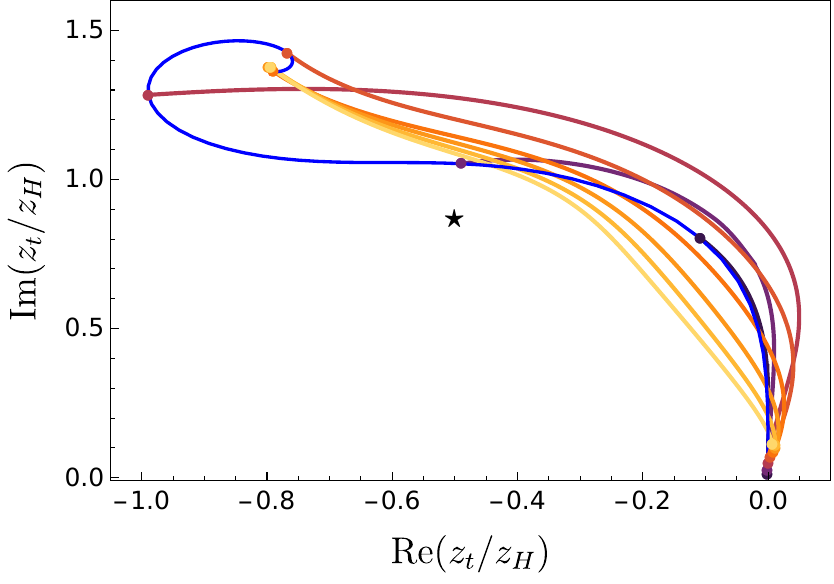}
     \caption{\textbf{Top panel}: complex $z_t$-plane trajectories of extremal surfaces associated with strips with $\Delta r/z_H = 1,2,4,6,8,10,12$ and $14$ and initial $\theta = \frac{\pi}{4} + 10^{-4}$, as $\theta$ varies along the path \eqref{alpha_complex}. \textbf{Bottom panel:} same as top panel, with $\theta$ going from $\frac{\pi}{4} + 10^{-4}$ to $\frac{\pi}{2}$. The upper vacuum-connected branch of complex extremal surfaces at $\theta = \frac{\pi}{2}$ is depicted in blue as in Fig.~\ref{fig:solution_space_Pi2}. The black star corresponds to a complex black hole horizon.}
     \label{fig:around_light cone}
\end{figure}

As an illustration of these results, in Fig.~\ref{fig:around_light cone} we show how the analytic continuation works in the $\varepsilon = 10^{{-}4}$ case. In the top panel, we plot the trajectories traced in the complex $z_t$-plane by the extremal surfaces with $\frac{\Delta r}{z_H} = 1,2,4,6,8$ and $10$, all smaller than $\frac{\Delta r_\textrm{min}(10^{-4})}{z_H} \approx 10.44$, as $\theta$ varies along the path \eqref{alpha_complex}. As it is manifest from the plot, the surface associated to the final timelike strip with $\theta = \frac{\pi}{4} + 10^{{-}4}$ is complex. The bottom panel of Fig.~\ref{fig:around_light cone} demonstrates that these complex surfaces, when $\theta$ is taken from $\frac{\pi}{4}+10^{-4}$ to $\frac{\pi}{2}$, always end up in the upper vacuum-connected branch of solutions shown in Fig.~\ref{fig:solution_space_Pi2}.

\vspace{1em}
\noindent \underline{$\Delta r \in [\Delta r_\textrm{min}(\varepsilon),\Delta r_\textrm{max}(\varepsilon)]$}.~In this case, as $\theta \to \frac{\pi}{4}^-$, several real extremal surfaces can contribute to the entanglement entropy. Therefore, to identify the complex extremal surfaces relevant to the timelike entanglement entropy computation, we must consider the analytic continuation of not only the vacuum-connected branch of real extremal surfaces, but also the unstable and horizon-connected ones. 

In this range of $\Delta r$, the analytic continuation of the real saddles in the vacuum-connected branch proceeds analogously to the $\Delta r < \Delta r_\textrm{min}(\epsilon)$ case. See the curves corresponding to $\frac{\Delta r}{z_H} = 12$ and $14$, all larger than $\frac{\Delta r_\textrm{min}(10^{-4})}{z_H}$ and smaller than $\frac{\Delta r_\textrm{max}(10^{-4})}{z_H}\approx14.17$, in Fig.~\ref{fig:around_light cone}. 

\begin{figure}[h!]
     \centering     \includegraphics[width=\linewidth]{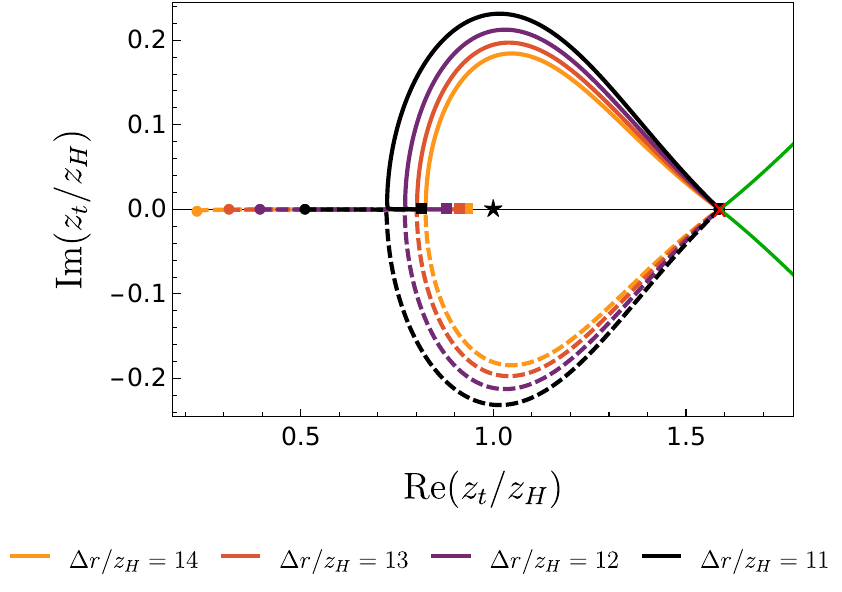}
     \caption{For $\Im(\theta)=-10^{-6}$, evolution in the complex $z_t$-plane of four solutions with $\Delta r/z_H = 11, 12, 13$ and $14$ in each of the complexified unstable (dashed curves) and horizon-connected branches (solid curves) as $\Re(\theta)$ ranges from $\frac{\pi}{4}-10^{-4}$ to $\frac{\pi}{2}$. Clearly, solutions in the complexified unstable (horizon-connected) branch end in the lower (upper) vacuum-disconnected branch.}   \label{fig:analytic_continuation_disconnected}
\end{figure}

On the other hand, we find that, if we analytically continue around the light cone as in Eq.~\eqref{alpha_complex} and then take $\theta \to \frac{\pi}{2}$, real saddles in both the unstable and horizon-connected branches flow to the vacuum-disconnected branches of complex saddles depicted in Fig.~\ref{fig:solution_space_Pi2}. This is our third main result in this section. A subtlety arises for a fixed $\Delta r$ as $\theta \to \frac{\pi}{2}$: at some intermediate angle $\theta_\star < \frac{\pi}{2}$, the solution originating from the unstable branch collides with the one from the horizon-connected branch. This collision makes it ambiguous to determine which solution connects to which vacuum-disconnected branch at $\theta = \frac{\pi}{2}$. To resolve this ambiguity, we introduce a small imaginary part to $\theta$ and define the branch assignment by taking the limit $\Im \theta \to 0$. Following this prescription, we find that solutions from the unstable (horizon-connected) branch flow to the lower (upper) vacuum-disconnected branch at $\theta = \frac{\pi}{2}$. See Fig.~\ref{fig:analytic_continuation_disconnected} for several examples of this behavior. Note that this explains why the vacuum-disconnected branches at $\theta = \frac{\pi}{2}$, originally found in Ref.~\cite{Heller:2024whi}, had to exist in the first place. 

\begin{figure}[h!]
     \centering     \includegraphics[width=\linewidth]{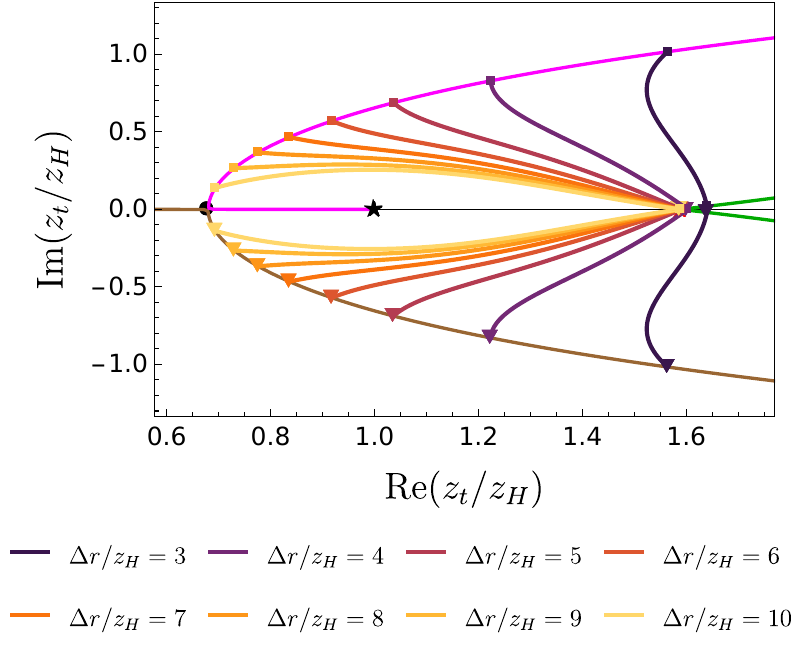}
     \caption{For $\Im(\theta) = -10^{-6}$, evolution of the complexified unstable and horizon-connected branches at $\Re(\theta) = \frac{\pi}{4}-10^{-4}$ (cf. Fig.~\ref{fig:solution_space_theta_complex}) as $\Re(\theta) \to \frac{\pi}{2}$. Clearly, the complexified unstable (horizon-connected)  branch maps to the lower (upper) vacuum-disconnected branch at $\Re(\theta)= \frac{\pi}{2}$, shown here for $\Im(\theta) = 0$ in green.}
     \label{fig:analytic_continuation_disconnected_2}
\end{figure}

Our analysis so far has established that, for $\Delta r \in [\Delta r_\textrm{min}(\varepsilon),\Delta r_\textrm{max}(\varepsilon)]$, the complex saddles in the vacuum-disconnected branches at $\theta = \frac{\pi}{2}$ descend from real saddles in the pre-light cone regime. It is natural to wonder where the remaining parts of these vacuum-disconnected branches come from. Given the results presented in Fig.~\ref{fig:solution_space_theta_complex}, a natural guess is that, for $\Delta r < \Delta r_\textrm{min}(\varepsilon)$, the vacuum-disconnected branches at $\theta = \frac{\pi}{2}$ descend from the pair of complex-conjugated branches of complex saddles emanating from $z_{t,\textrm{min}}(\varepsilon)$. This expectation is confirmed by the results shown in  Fig.~\ref{fig:analytic_continuation_disconnected_2}.

\subsection{Area densities and timelike entanglement entropy}
\label{sec:black_brane_TEE}

\begin{figure}[h!]
\begin{center}
\includegraphics[width=\linewidth]{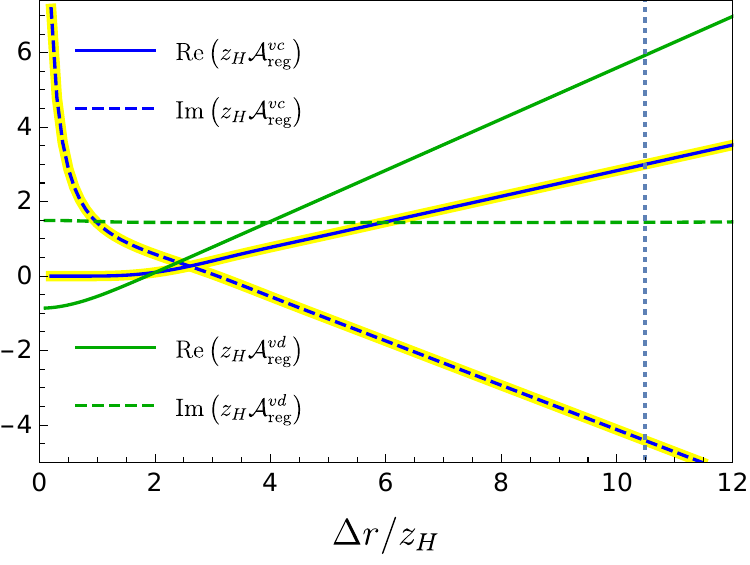}
\caption{\small For a timelike strip with $\theta = \frac{\pi}{2}$, regularized area density $\mathcal{A}_\textrm{reg}$ for the upper vacuum-connected (blue curves) and upper vacuum-disconnected (green curves) branches of extremal surfaces. Real (imaginary) parts correspond to solid (dashed) curves. The dotted vertical line marks the location of $\Delta r_\textrm{min}^\star$, with the vacuum-disconnected branches becoming available saddles to its right. The dominant contribution to the timelike entanglement entropy according to our prescription has been highlighted in yellow.} 
\label{fig:A_Pi2}
\end{center}
\end{figure}

We are finally in a position to employ our prescription to compute the timelike entanglement entropy. We begin by examining the case $\theta = \frac{\pi}{2}$. The complex extremal surfaces are shown in Fig.~\ref{fig:solution_space_Pi2} and their corresponding area densities in Fig.~\ref{fig:A_Pi2}. 

To evaluate the timelike entanglement entropy, we must consider two distinct regimes:
\begin{itemize}
\item First, for $\Delta r < \Delta r_\textrm{min}^\star$, our analysis from the previous subsection shows that only the complex extremal surfaces in the upper vacuum-connected branch contribute, and hence the timelike entanglement entropy computed according to our prescription upholds the UV--IR correspondence by construction. 

\item Second, for $\Delta r \geq \Delta r_\textrm{min}^\star$, the vacuum-disconnected branches of complex extremal surfaces also become potential contributors, in addition to the upper vacuum-connected branch. We must choose the solution with the smallest real part of the area density. As shown in Fig.~\ref{fig:A_Pi2}, in this regime the vacuum-disconnected branches always have a larger $\Re \mathcal A_\textrm{reg}$ than the vacuum-connected branch. As a result, the upper vacuum-connected branch continues to dominate.
\end{itemize}
In conclusion, according to our prescription, the timelike entanglement entropy for a strip with $\theta = \frac{\pi}{2}$ is always determined by the upper vacuum-connected branch of complex extremal surfaces. 
\begin{figure}[h!]
\begin{center}
\includegraphics[width=\linewidth]{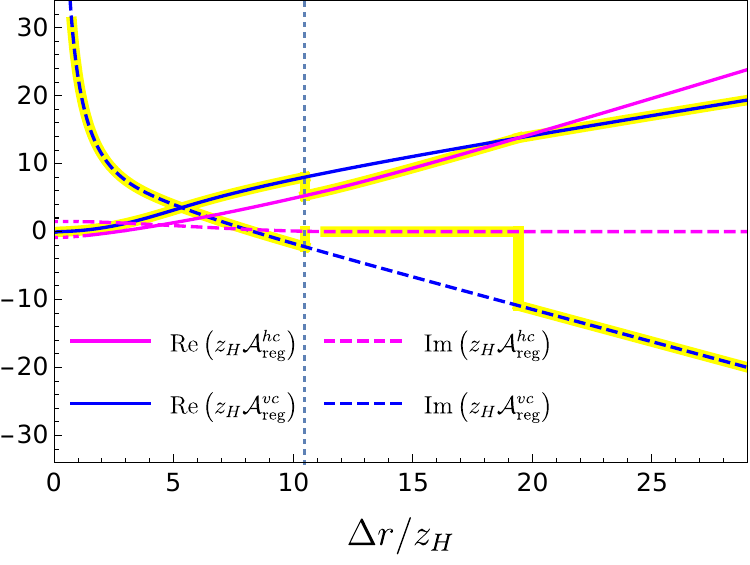}
\caption{\small For a timelike strip with $\theta = \frac{\pi}{4} + 0.002$, regularized area density $\mathcal{A}_\textrm{reg}$ of the analytical continuation of the vacuum-disconnected (blue) and horizon-connected (magenta) branches ($\mathcal{A}_\textrm{reg}$ for the analytical continuation of the unstable branch is the complex conjugate of the horizon-connected one and not shown). Real (imaginary) parts emerging from our numerical computation correspond to solid (dashed) curves, while dotted curves denote that the corresponding quantity has been obtained through a numerical extrapolation. The dotted vertical line marks the location of $\Delta r_\textrm{min}^\star$, and the dominant contribution to the timelike entanglement entropy according to our prescription have been highlighted in yellow.  Interestingly, in this case minimization of $\Re \mathcal{A}_\textrm{reg}$ together with connectedness to holographic entanglement entropy induces a jump in $\Re \mathcal{A}_\textrm{reg}$.} 
\label{fig:A_near-lightcone}
\end{center}
\end{figure}

The computation of the timelike entanglement entropy for $\theta \in \left(\frac{\pi}{4}, \frac{\pi}{2}\right)$ proceeds in an analogous way and the same conclusion follows, provided that for $\Delta r > \Delta r_\textrm{min}^\star$ the area density of the vacuum-connected branch is lower than the rest. Empirically, we find that this is always the case except when $\theta$ is sufficiently close to $\frac{\pi}{4}$.  In this immediate vicinity of the light cone, the vacuum-disconnected branch has smaller $\Re \mathcal{A}_\textrm{reg}$ that the vacuum-connected ones at $\Delta r = \Delta r_\textrm{min}^\star$ and, as a consequence, the timelike entanglement entropy features a zeroth-order phase transition as soon as the vacuum-disconnected complex extremal surfaces become available saddles. This zeroth-order phase transition gives way to a first-order phase transition at larger separations, where the vacuum-connected branch becomes dominant again. See Fig.~\ref{fig:A_near-lightcone} for an example.

We conclude this section with two comments. The first is that, if we were to compute the timelike entanglement entropy by minimizing over \emph{all} available complex extremal surfaces, we would find that, for a fixed $\theta$, the vacuum-disconnected solutions dominate at sufficiently small $\Delta r$. Our analysis further shows that these vacuum-disconnected solutions for $\theta > \frac{\pi}{4}$ originate from complex saddles at $\theta < \frac{\pi}{4}$. Thus, the requirement that the timelike entanglement entropy preserves the UV–IR correspondence suggests that complex extremal surfaces should never be treated as contributing subleading saddles in holographic entanglement entropy computations.

The second and final comment is that, given the presence of the zeroth-order phase transition in the timelike entanglement entropy in the immediate vicinity of the light cone, the reader might rightfully wonder why we have not decided to simply define the timelike entanglement entropy through the analytical continuation of the vacuum-connected branch of real extremal surfaces which, for a given $\Delta r$, always gives the dominant contribution to the entanglement entropy infinitesimally before the light cone. In particular, note that this choice would also naturally uphold the UV--IR correspondence and, in addition, lead to a smooth answer for all angles. Our main reason for not pursuing this alternative definition is that in two-dimensional conformal field theories---where entanglement entropy is derived from a two-point correlator of twist operators---we expect the timelike entanglement entropy---defined via analytic continuation of this correlator---to become singular whenever the insertion points are null-separated. As we will show in the next section, our prescription naturally reproduces these null singularities, whereas the naive analytical continuation of the saddle that dominates immediately before the light cone does not.

\section{Holographic thermal state on~$\mathbb{R}\times S^{1}$ \label{sec.RS1}}

We will now show that also in two-dimensional holographic conformal field theories with a compact spatial direction the prescription advocated in the present paper and outlined in Sec.~\ref{sec:idea} gives a physically sensible result, as it correctly identifies the presence of light cone singularities when the endpoints of the entangling region are null-separated. Indeed, in two-dimensional conformal field theories the entanglement entropy of a segment $a$ is defined in terms of a two-point correlator of twist operators $\sigma_n$, $\bar\sigma_n$ evaluated at its endpoints, which implement the appropriate boundary conditions in the replica manifold
\cite{Calabrese:2004eu},
\begin{equation}\label{eeTwist}
S_a=\lim_{n\to 1} \frac{1}{1-n}\log{\langle\sigma_n \bar\sigma_n\rangle}.
\end{equation}
For a two-dimensional conformal field theory with a compact spatial direction $\phi$ in the vacuum state, Eq.~\eqref{eeTwist} leads to
\begin{equation}\label{eeCFT2}
S_a=\frac{1}{4 G_{N}}\log{\left(\frac{4}{\delta^2}\sin{\frac{\Delta t+\Delta \phi}{2}}\sin{\frac{\Delta t-\Delta \phi}{2}}\right)}\,,
\end{equation}
where $\Delta t$ and $\Delta \phi$ are the coordinate differences between the endpoints of the entangling region and $\delta\ll 1$ is again a UV regulator \cite{Calabrese:2004eu,Doi:2023zaf}. The entropy \eqref{eeCFT2} is singular when the two endpoints are null-separated, i.e.\ $\Delta t=\Delta \phi$, but also $\Delta t=2\pi-\Delta \phi$. The latter condition arises due to the compact nature of the spatial direction and can be understood from the fact that, on a fixed spacetime slice, any pair of points defines two intervals on the surface of the Lorentzian cylinder $\mathbb R \times S^1$, depending on the choice of leaving an endpoint clockwise or anti-clockwise. As these singularities arise purely from geometrical properties of the boundary spacetime and do not depend on the bulk geometry, we will refer to them as \emph{kinematical singularities}.

For this same reason, we expect such kinematical singularities to arise in thermal and excited states, as well. In these cases, the entanglement entropy for an interval cannot be expressed in closed form from the conformal field theory side. Still, it can be evaluated on the gravity side, where multiple competing configurations arise due to the presence of a nontrivial topological structure in the bulk (conical defect or black hole horizon). The goal of this section is to show that such divergences do arise holographically only if the minimization over competing configurations, according to the prescription outlined in Sec.~\ref{sec:idea}, is applied \emph{after} the analytic continuation through the light cone. Hence, this is a further example showing that the two operations, namely selection of the dominant saddle by minimization and analytical continuation, do \emph{not} commute in general, emphasizing the need for the advocated prescription, which specifies their ordering.

\subsection{Setup}
We consider the family of three-dimensional bulk metrics parametrized by the real parameter $\mu\geq -1$,
\begin{equation}\label{AdS3_metric}
ds^2=-\left(\rho^2-\mu\right) dt^2+\frac{d\rho^2}{\rho^2-\mu}+\rho^2d\phi^2\,,
\end{equation}
where $\rho\geq0$, $\phi\in[0,2\pi)$ and the asymptotic boundary where the dual conformal field theory lives is at $\rho\to\infty$. Comparing with Eq.~\eqref{metric}, $\rho$ plays the role of $1/z$ and we use $\rho$ as it is more convenient in the present setup.

All the geometries encapsulated by Eq.~\eqref{AdS3_metric} represent the time development of states of two-dimensional conformal field theories on the Lorentzian cylinder $\mathbb{R}\times S^{1}$ of a unit radius. When $\mu=-1$, the line element \eqref{AdS3_metric} represents the gravity dual to the vacuum state. If $-1<\mu<0$, it represents global AdS$_3$ with the insertion of a conical defect at $\rho = 0$, which corresponds to an excited state. Finally, if $\mu> 0$, the geometry~\eqref{AdS3_metric} describes the BTZ black hole~\cite{Banados:1992wn}, dual to the thermal state with temperature~$T=\sqrt{\mu}/(2\pi)$.

\subsection{Entanglement entropy on Cauchy slices}

\begin{figure}[h!]
    \centering    \includegraphics[width=0.95\linewidth]{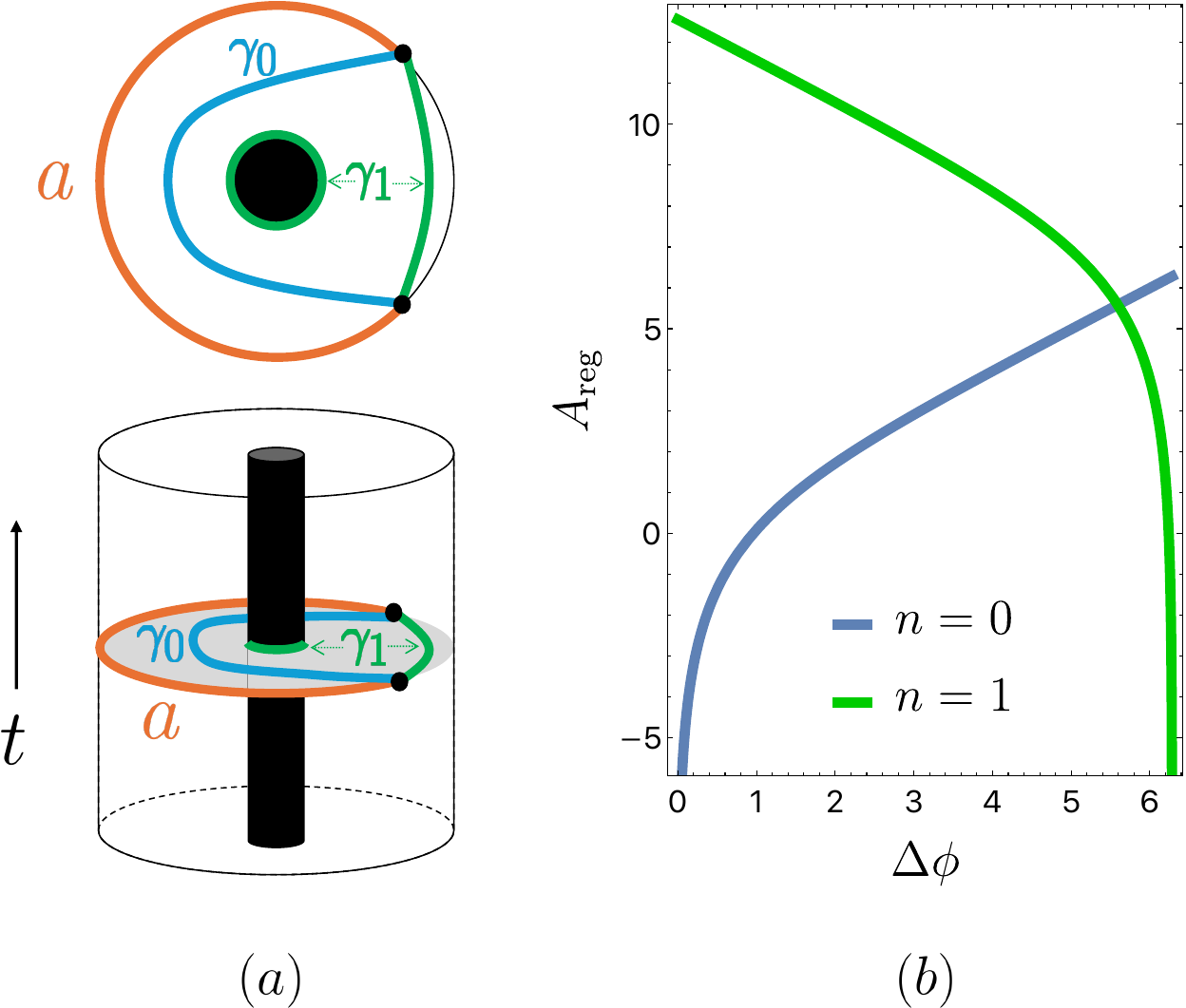}
    \caption{\textbf{(a)} Sketch of two possible geodesics arising in the BTZ black hole metric given by~\eqref{AdS3_metric} with $\mu>0$, when the boundary interval $a$ lies on a constant time slice: $\gamma_0$ (blue) with winding number $n=0$, and $\gamma_1$ (green) with $n=1$. Note that the latter includes an additional contribution around the black hole horizon to respect the homology constraint. Geodesics with a higher number of windings (not shown in the picture) also exist. \textbf{(b)} Regularized length $A_{\mathrm{reg}}\equiv A+2\log{\delta}$ for the two geodesic configurations. While for small separations $\gamma_0$ dominates, there exists a value $\Delta \phi_*<2\pi$ at which the two saddles exchange dominance.}
    \label{fig:wrappinggeodesics}
\end{figure}

The entangling region of interest for this section is a single interval specified by a pair of boundary spacelike-separated points offset by $\Delta t$ in time and $\Delta \phi$ in the angular direction. The entanglement entropy is then computed in terms of bulk geodesics connecting these points. For $\mu > -1$, i.e.\ in excited or thermal states, multiple geodesics exist that are enumerated by their winding number $n\geq 0$ around the conical defect at $\rho=0$ (if $-1<\mu<0$) or the black hole horizon (if $\mu>0$). All these different configurations should be considered as potential duals to the entanglement entropy. The geodesic connecting the region endpoints with winding number $n=0$ has length
\begin{align}
A= &\log \left[\frac{2}{\mu \delta^2} \left(\cosh{\Delta \phi\sqrt{\mu}}-\cosh{\Delta t \sqrt{\mu}}\right)\right],\label{BTZdist}
\end{align}
where we suppressed terms with positive powers of the UV regulator $\delta$. When $-1\leq \mu<0$ the hyperbolic functions become trigonometric ones, hence the entropy has a periodicity which depends on the defect mass and coincides with the spatial one only in the vacuum $\mu=-1$. When $\mu>0$ we have instead the typical linear behavior at large separations which is expected in a thermal state. A sketch of such a geodesic in the case $\Delta t=0$ is $\gamma_0$ in Fig.~\ref{fig:wrappinggeodesics}(a). A geodesic with generic winding number $n$ has length given by \eqref{BTZdist} with $\Delta \phi\to 2\pi n-\Delta \phi$. Compare with $\gamma_1$ in Fig.~\ref{fig:wrappinggeodesics}(a) for a pictorial representation of the case with a single winding~($n=1$).

If $\Delta t=0$, i.e.\ the two points lie on the same constant-time slice, the holographic entanglement entropy follows a well-known behavior: there exists a critical separation $\Delta \phi=\Delta \phi_*$ at which a phase transition occurs between the configuration $\gamma_0$ that does not wrap the defect (or the horizon) and $\gamma_1$ that instead does once. When $\Delta \phi>\Delta\phi_*$, the length of $\gamma_1$ becomes smaller than that of $\gamma_0$, see Fig.~\ref{fig:wrappinggeodesics}(b), and hence $\gamma_1$ dominates. Note that only the geodesics with the two lowest winding numbers $n=0,1$ are the ones that exchange dominance under the criterion of selecting the one with minimal length \cite{Balasubramanian:2014sra}.

When there is a black hole horizon in this setup, the homology constraint plays a crucial role. The homology constraint consists of the requirement that there exists a codimension-one interpolating homology surface whose only boundaries are the entangling surface $\gamma$ and the boundary subregion $a$. It is usually motivated by the fact that the causal wedge of the boundary subregion has to be contained within its entanglement wedge \cite{Hubeny:2013gba}. As a consequence, when $\mu>0$ the wrapping configuration $\gamma_1$ includes a contribution coming from a disconnected piece encircling the horizon, see Fig.~\ref{fig:wrappinggeodesics}(a).

These considerations apply equally to intervals that lie on tilted spacetime slices, i.e.\ whose endpoints are separated by $\Delta \phi$ along the spatial direction and $\Delta t$ along time, provided that the two endpoints are spacelike separated
\begin{equation}
\label{eq.condspat}
\Delta t<\min{(\Delta \phi,2\pi-\Delta\phi)},
\end{equation}
where the term $2\pi -\Delta \phi$ originates from the compactness of the spatial circle giving rise to two ways of connecting a pair of points. Note that the condition~\eqref{eq.condspat} can be understood as the condition for the existence of a global spatial (time) slice containing the subregion, or, in other words, demanding that the complement is also spatial. We will be interested in such more general tilted subsystems, as they are intermediate steps in the analytic continuations being part of our prescription for holographic timelike entanglement entropy.

When the condition~\eqref{eq.condspat} is satisfied, the bulk extremal surfaces (here: geodesics) are real and hence potential contributors to the standard entanglement entropy. The homology constraint is also understood in the same manner as on the constant-time slice considered above.

When the inequality~\eqref{eq.condspat} saturates, then either a subregion or its complement become null. This is a singular limit for entanglement entropy and we will analytically continue across it using our prescription, see Fig.~\ref{fig:rotation}. 

\subsection{Analytic continuation past the light cone}

In the following we will employ the prescription outlined in~Sec.~\ref{sec:idea} in the three dimensional bulk setup~\eqref{AdS3_metric}. We will utilize
\begin{equation}\label{rotationGlobal}
\Delta \phi = \Delta r \cos{\theta} \quad \mathrm{and} \quad \Delta t = \Delta r \sin{\theta}.
\end{equation}
There will be three key differences with respect to the case studied in Sec.~\ref{sec.R12}, both originating from the compactness of the spatial direction:
\begin{itemize}
\item The maximal value of $\Delta r$ that we can take starting with entanglement entropy is naturally limited by~$2\pi$. As a result, the outcome of the prescription outlined in Sec.~\ref{sec:idea} at any value of the `rotation' angle, in particular at its maximal value $\theta = \frac{\pi}{2}$, will be limited to $\Delta r= 2 \pi$. While in the tensor network picture of temporal entanglement there does not seem to be a need for such a limitation, in this section will just accept it as a feature of the prescription relying on the analytic continuation depicted in Fig.~\ref{fig:rotation} and in Sec.~\ref{sec.outlook} we will speculate on how to beyond $\Delta r = 2 \pi$. 
\begin{figure}
    \centering   \includegraphics[width=0.6\linewidth]{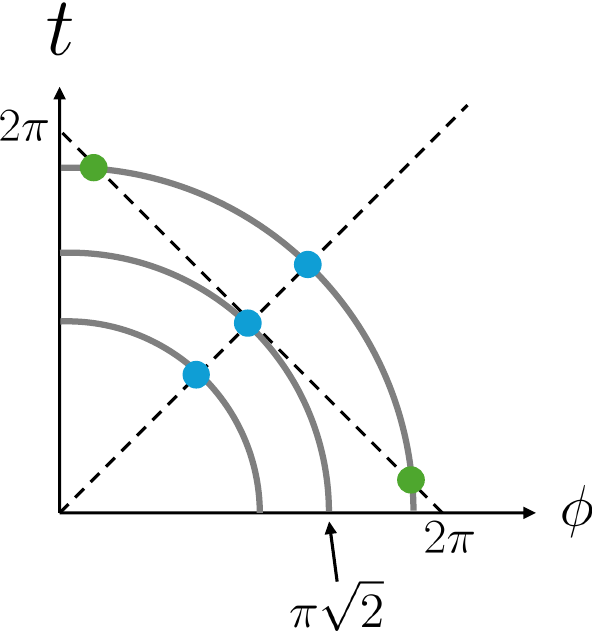}
    \caption{Emergence of null singularities along the rotation \eqref{rotationGlobal} as a function of the boundary interval size $\Delta r$. There is a critical value $\Delta r=\pi\sqrt{2}$ below which only the singularity at $\theta=\frac{\pi}{4}$ appears in the configuration with $n=0$ (blue dots). For $\Delta r>\pi\sqrt{2}$, instead, two more kinematical singularities arise when the endpoints are null-separated by hitting the light cone emitted by the other endpoint (green dots, compare with Figs.~\ref{fig:continuationvacuum},~\ref{fig:continuationdefect},~\ref{fig:continuationBTZ} for a plot of the divergences).}
    \label{fig:light conesing}
\end{figure}
\item The compact nature of the spatial direction leads to the appearance of null singularities not only at $\theta = \frac{\pi}{4}$, but also at the other values of $\theta$ when the points become connected via a null geodesics going across the other side of the cylinder, see Fig.~\ref{fig:light conesing}. Which one is hit first depends on the value of $\Delta r$:
\begin{itemize}
\item[\#] $\Delta r < \frac{1}{\sqrt{2}} \times 2 \pi$: in this case the `rotation' encounters only the familiar light cone singularity when the subregion itself becomes null at $\theta = \frac{\pi
}{4}$. This is avoided by a small detour into the complex $\theta$ plane as in Fig.~\ref{fig:rotation}. The subsequent `rotation' to the timelike regime of $\frac{\pi}{4}<\theta \leq \frac{\pi
}{2}$ does not encounter any additional singularities as the complement remains spacelike, compare with Fig.~\ref{fig:light conesing}.
\item[\#] $\frac{1}{\sqrt{2}} \times 2 \pi < \Delta r < 2 \pi$: full `rotation' to $\theta = \frac{\pi}{2}$ encounters three null singularities, see Fig.~\ref{fig:light conesing}. The first one $\theta_1$ is associated with the complement becoming null and only for $\theta$ smaller than this threshold value the notion of entanglement entropy still applies. The second null singularity is associated with the subregion itself becoming null and is still at $\theta_2=\frac{\pi}{4}$. The third one $\theta_3$ is associated with complement becoming null again. The kinematical singularities arising from the complement becoming null do not depend on the state and are given by
\begin{equation}\label{kinematicsing}
\theta_{1,\, 3}=\arctan{\frac{2 \pi \mp\sqrt{2} \sqrt{\Delta r^2-2 \pi ^2}}{2\pi \pm \sqrt{2} \sqrt{\Delta r^2-2 \pi ^2}}}\,.\hspace{-3em}
\end{equation}
All these null singularities are dealt with by the excursion onto the complex $\theta$ plane, as in Fig.~\ref{fig:rotation}, but at the respective real values of the angle.
\end{itemize}
\item Whereas in Sec.~\ref{sec.R12} all real extremal surfaces in the spatial regime satisfied the homology constraint, this is not longer the case here for $\mu > 0$ and has to be taken into account before the analytic continuation past the first light cone takes place.
\end{itemize}
Below we will go between three classes of solutions one-by-one and discuss the emerging picture for the holographic timelike entanglement entropy. We will always consider the regime $\Delta r>\pi\sqrt{2}$, where all the kinematical singularities arise, as they will be a crucial element in testing our prescription.

\begin{figure}
    \centering
    \includegraphics[width=0.95\linewidth]{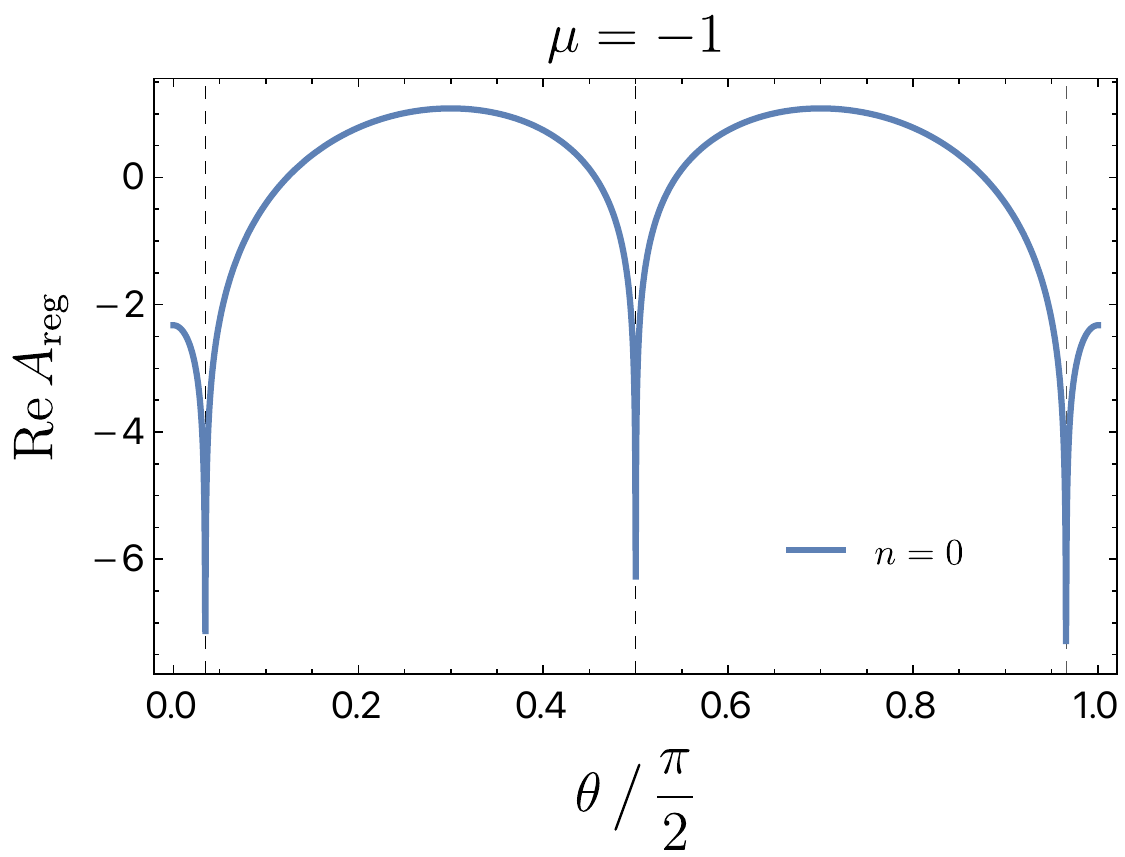} 
    
    \vspace{1em}
    \includegraphics[width=0.95\linewidth]{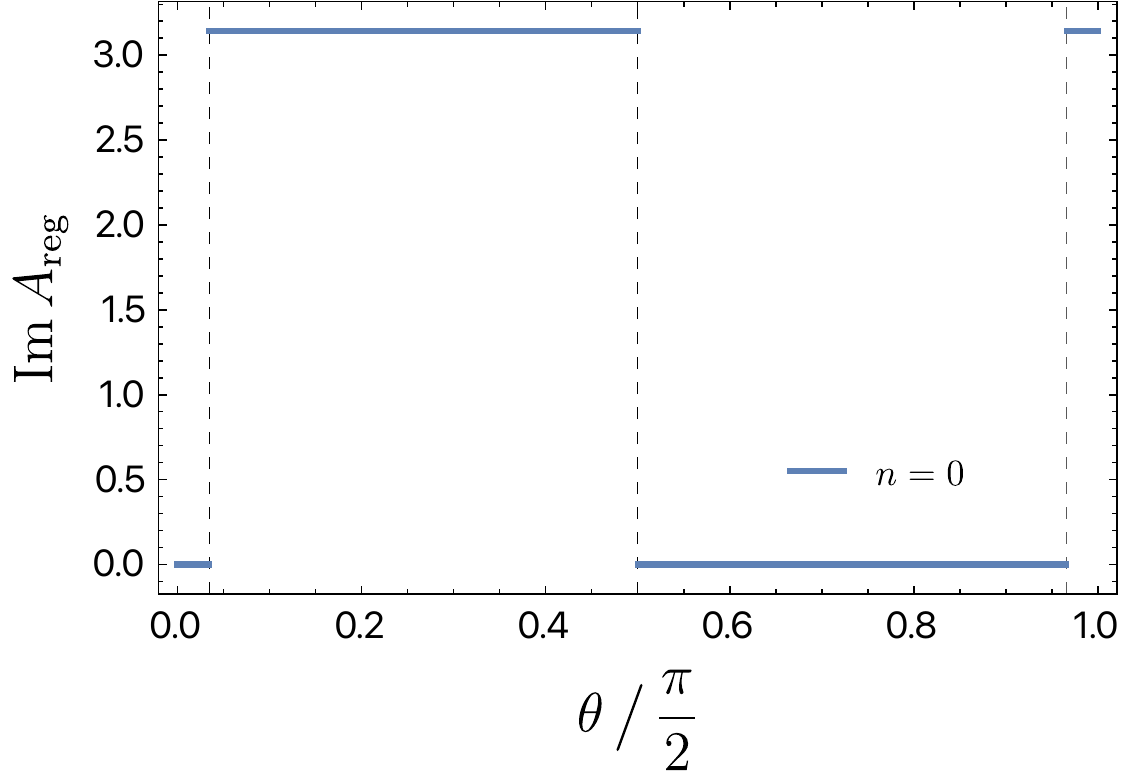} 
       
    \caption{Analytic continuation of holographic entanglement entropy in the vacuum state, $\mu=-1$. The size of the boundary interval $\Delta r=1.9\pi$ is fixed to be larger than the critical value $\pi\sqrt{2}$ past which additional kinematical singularities appear, and we vary $\theta$ from 0 (constant time slice) to $\frac{\pi}{2}$. Only the configuration $n=0$ contributes in this case. \textbf{Top panel}: Real part, which exhibits the kinematical singularities. \textbf{Bottom panel}: Imaginary part plotted modulo $2\pi$, which assumes the values $0$ or $\pi$ depending on the endpoints being spacelike or timelike separated.}
    \label{fig:continuationvacuum} 
\end{figure}

\begin{itemize}
\item \textbf{Vacuum ($\mu = -1$):} In the case of the empty AdS$_{3}$ geometry, there is only one extremal surface for each pair of points specifying the subregion and its complement. When both the subregion and its complement are spatial, the surface is real and trivially satisfies the homology constraint. The analytic continuation past each singularity is therefore unique and does not carry any ambiguity. See Fig.~\ref{fig:continuationvacuum} for an example.
\end{itemize}

\begin{figure}
    \centering
    \includegraphics[width=0.95\linewidth]{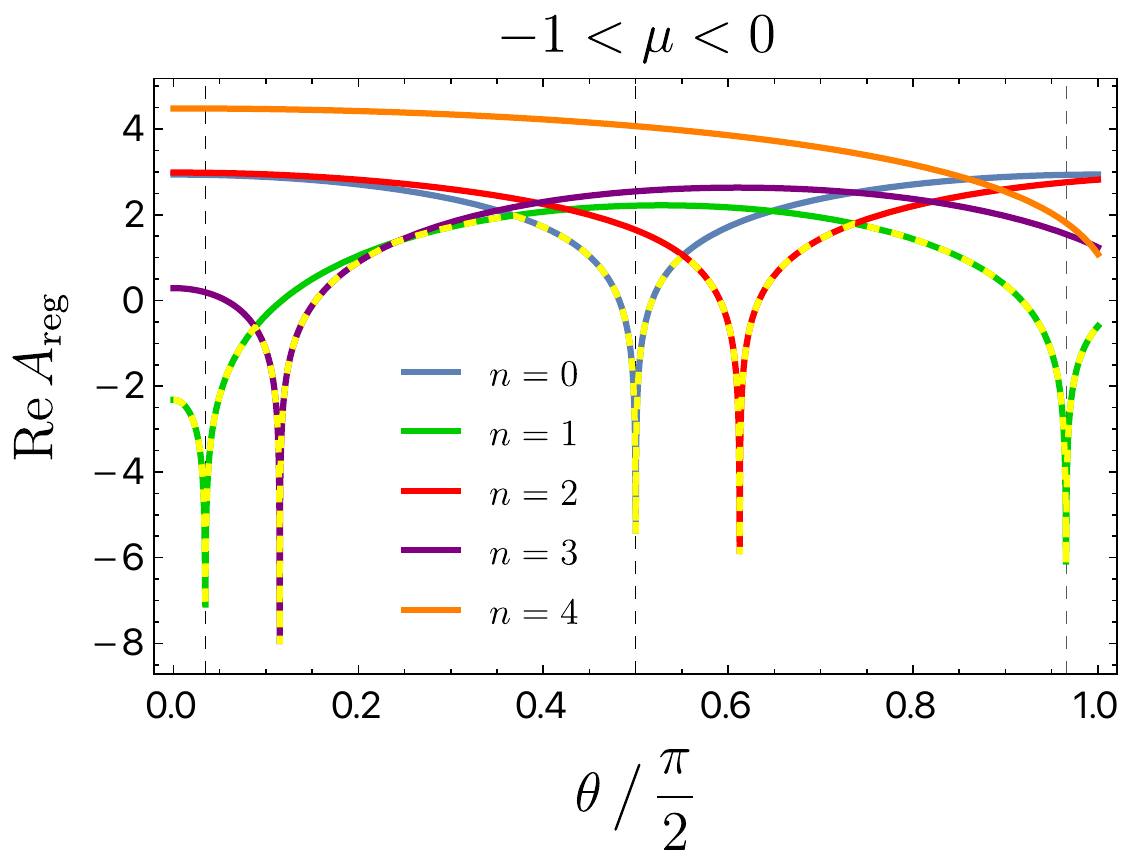} 

    \vspace{1em}
    \includegraphics[width=0.95\linewidth]{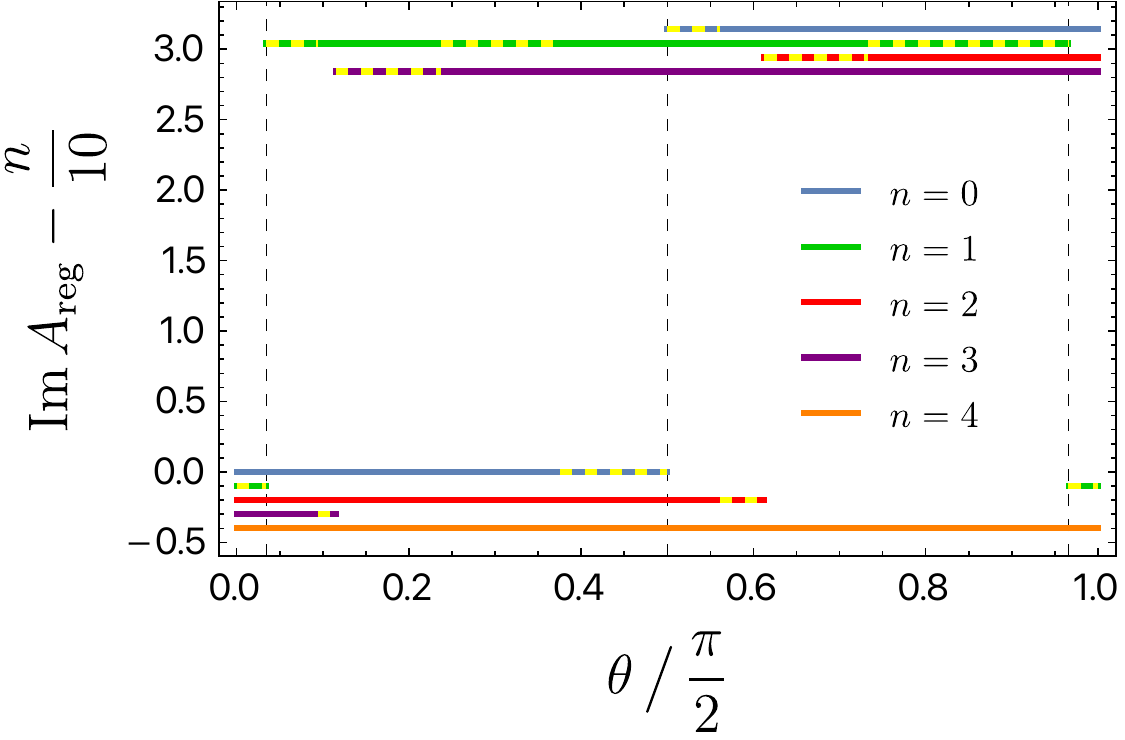} 
       
    \caption{Analytic continuation of holographic entanglement entropy in the excited state dual to AdS$_3$ with insertion of a conical defect, $-1<\mu<0$. The size of the boundary interval is $\Delta r=1.9\pi$, as in Fig.~\ref{fig:continuationvacuum}. Depending on the value of the defect mass $\mu$ additional bulk singularities arise on top of the kinematical ones that were already apparent in the vacuum state of Fig.~\ref{fig:continuationvacuum}. The value of $\mu$ determines also which configurations contribute to the entropy: in this case, we chose the value $\mu=-0.2$ to avoid clutter and geodesics with $n\leq 3$ are relevant. The dominant configurations selected by minimization are highlighted in yellow. \textbf{Top panel}: Real part. \textbf{Bottom panel}: Imaginary part plotted modulo $2\pi$, which still assumes the two discrete values 0, $\pi$. An $n$-dependent shift has been introduced for readability.}
    \label{fig:continuationdefect} 
\end{figure}

\begin{itemize}
\item \textbf{Conical defect ($-1<\mu<0$):} In this case, multiple configurations arise depending on their winding number $n$ around the conical defect. They feature not only the kinematical singularities \eqref{kinematicsing}, but also \emph{bulk singularities} associated with null connectivity over the bulk, see, e.g.,~\cite{Hubeny:2006yu,Gary:2009ae,Maldacena:2015iua,Dodelson:2023nnr} for corresponding discussions in the context of boundary correlation functions. Bulk singularities depend on the structure of the dual spacetime, in this case on the mass of the defect $\mu$. As it can be seen from Fig.~\ref{fig:continuationdefect}, bulk singularities arise up to a given winding number which also depends on the value of $\mu$. The homology constraint is again automatically satisfied, as the minimal contour encircling the defect has zero measure. However, the analytic continuation is no more unique, as when crossing each singularity as described in Sec.~\ref{sec:idea} there is an ordering ambiguity between crossing the light cone and selecting the configuration with minimal real part. This issue will be addressed at the end of this section, showing that the prescription described in Sec.~\ref{sec:idea} solves this ambiguity coherently with field theory expectations. 
\end{itemize}

\begin{figure}
    \centering
    \includegraphics[width=0.95\linewidth]{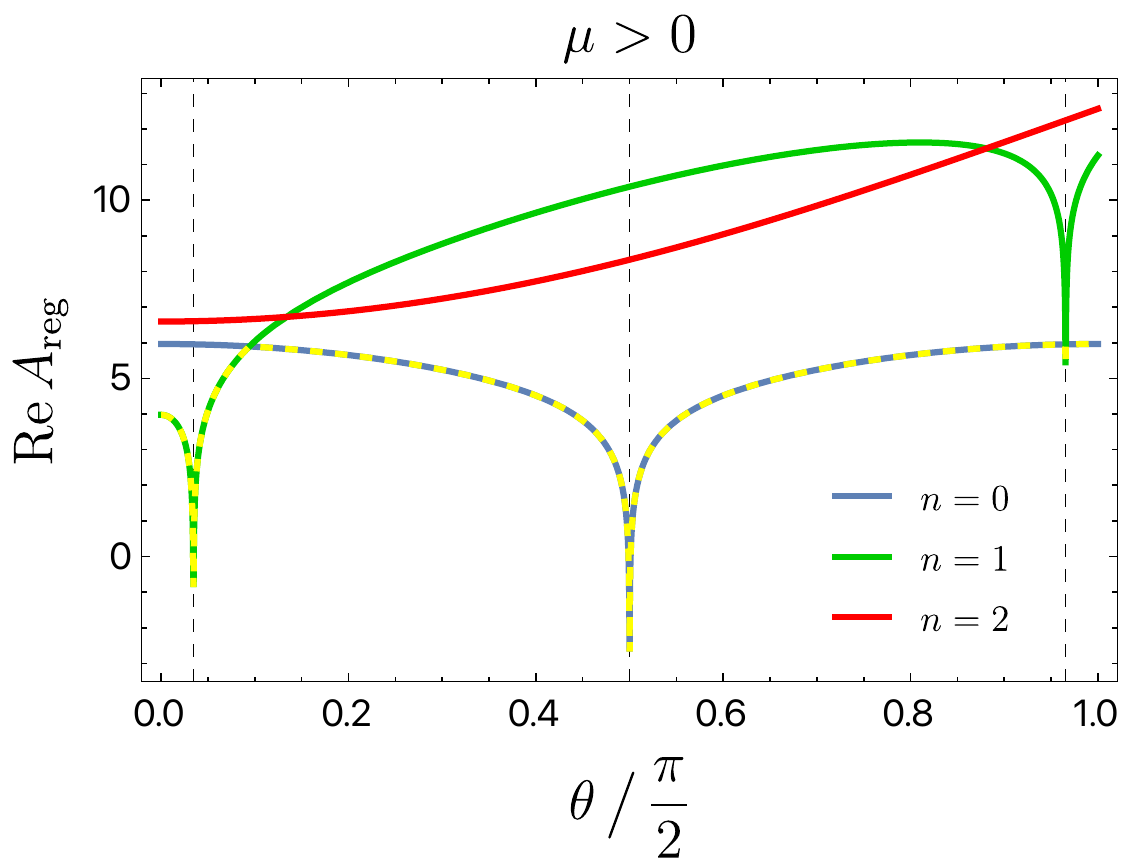} 

    \vspace{1em}
    \includegraphics[width=0.95\linewidth]{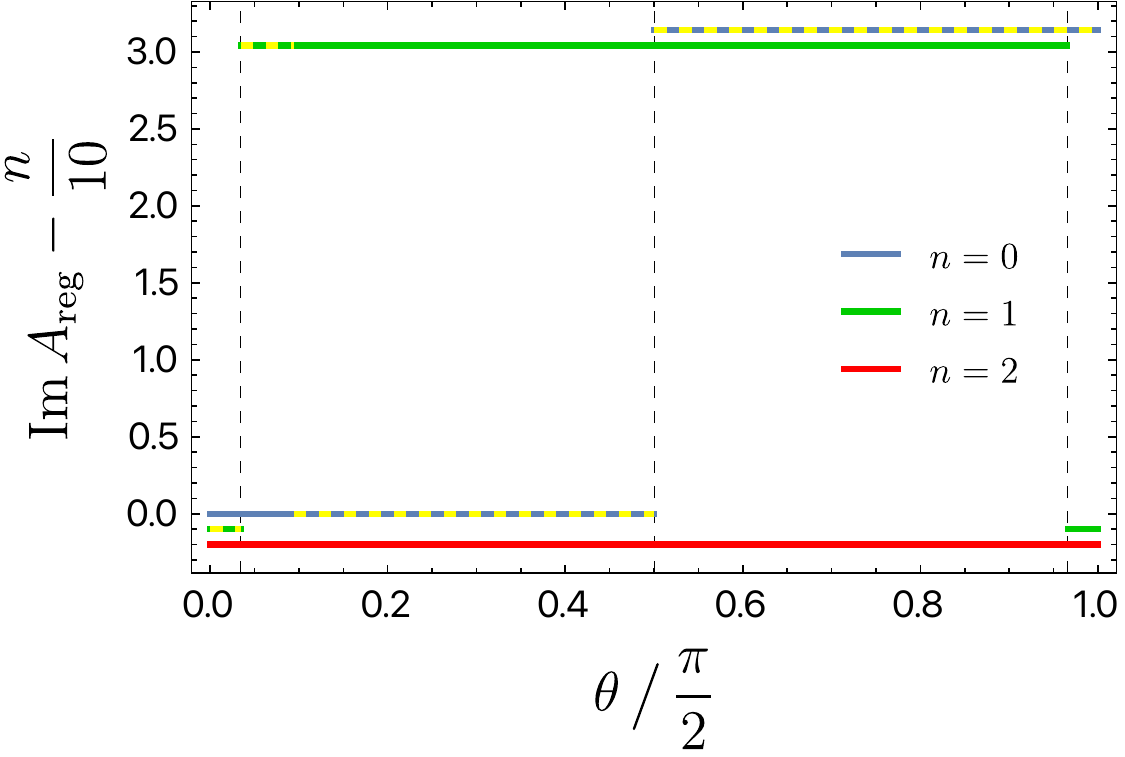} 
       
    \caption{Analytic continuation of holographic entanglement entropy in the thermal state dual to the BTZ black hole, $\mu>0$. The size of the boundary interval is $\Delta r=1.9\pi$, as in Figs.~\ref{fig:continuationvacuum},~\ref{fig:continuationdefect} and $\mu=1$. In this case, only geodesics with $n<2$ contribute to the entropy. The homology constraint is enforced for the configuration $n=1$ by adding the horizon length $2\pi\sqrt{\mu}$. The dominant configurations are highlighted in yellow. \textbf{Top panel}: Real part. \textbf{Bottom panel}: Imaginary part plotted modulo $2\pi$, which still assumes the two discrete values 0, $\pi$. An $n$-dependent shift has been introduced for readability.}
    \label{fig:continuationBTZ} 
\end{figure}

\begin{itemize}
\item \textbf{BTZ black hole ($\mu>0$):} In this case, as expected, the same kinematical singularities \eqref{kinematicsing} arise, see Fig.~\ref{fig:continuationBTZ}. An additional complication with respect to the conical defect case is the homology constraint, which now has to be carefully enforced at any value of $\theta$. As discussed before, on constant time slices the contribution from the homology constraint is typically taken to be the length of a curve wrapping the horizon, i.e.\ $2\pi\sqrt{\mu}$, for any geodesic which wraps the horizon an odd number $n$ of times. Intuitively, this condition emerges from the fact that a geodesic with $n$ even has always a piece around the horizon which acts as a ``screen'' for the others, ensuring the existence of an interpolating surface between the extremal surface and the boundary subregion required to satisfy the homology constraint. This is not the case for $n$ odd, as the innermost part of the curve will always need a further screen from an additional piece encircling the horizon, see Fig.~\ref{fig:wrappinggeodesics}.
\end{itemize}

We will now show how our prescription deals with the kinematical singularities, focusing on the last two cases (conical defect and BTZ black hole, $\mu>-1$), where ambiguities in the analytical continuation arise due to the existence of multiple extremal surfaces (here: geodesics). Consider the first kinematical singularity $\theta_1$. According to the prescription of Sec.~\ref{sec:idea}, the entropy for $\theta>\theta_1$ is defined by analytical continuation of all the real geodesic configurations at $\theta=0$ up to $\theta$, and then by selecting the configuration with the smallest real part. This means that the minimization at $\theta$ occurs \emph{after} crossing the light cone at $\theta_1$. Keeping only the saddle with minimal real part \emph{before} crossing the singularity at $\theta_1$, instead, does not allow to reproduce the next singularities $\theta_i>\theta_1$, as they can arise from saddles which are not dominating at $\theta_1$. This happens whenever there is an exchange of dominance between different saddles after the singularity $\theta_1$, which is often the case: compare with Figs.~\ref{fig:continuationvacuum},~\ref{fig:continuationdefect}. The same argument applies for all the other singularities~$\theta_i$.

As a concrete example, consider again Figs.~\ref{fig:continuationvacuum},~\ref{fig:continuationdefect} and focus on the kinematic singularities. An exchange of dominance arises between the configurations $n=0$ and $n=1$ as a function of $\theta$. Let us consider for instance the singularity $\theta_2=\frac{\pi}{4}$ (any other would lead to the same result). Even if for $\theta=0$ the configuration with winding number $n=1$ dominates (this is the case both for the conical defect and the BTZ black hole), as the light cone is approached there is a phase transition such that when $\theta\to\frac{\pi}{4}^-$ the dominant configuration will always be the one with $n=0$. Indeed, in this limit the lengths \eqref{BTZdist} are approximated by
\begin{align}
A= &\log{\left[\frac{2}{\mu\delta^2}\left( \cosh{\sqrt{\mu}\left(2n\pi-\frac{\Delta r}{\sqrt{2}}\right)}-\cosh{\Delta r\sqrt{\frac{\mu}{2}}} \right)\right]}\nonumber\\[.5em]
& +\mathcal{O}\left(\theta-\frac{\pi}{4}\right).\label{global_light cone}
\end{align}
Clearly, in the $\theta \to \frac{\pi}{4}^-$ limit the dominant configuration for any value of $\Delta r$ is given by the geodesic with no windings $n=0$, as the length above diverges to $-\infty$ while the others remain finite. From this kind of limits it is always possible to show that when a configuration with given $n$ diverges the others remain finite, hence minimization \emph{after} crossing the light cone will always pick the one exhibiting the kinematical divergence. Finally, note that there can be light cone divergences also for fine-tuned values of $\Delta r=n\pi\sqrt{2}$ such that the $n$-windings configuration is equally divergent to $-\infty$. This does not affect the statement that, in general, the dominant configuration does not correspond to the one leading to the expected light cone singularities in the timelike regime.

As a consequence, if timelike entanglement entropy is defined through the analytic continuation of the saddle giving the dominant contribution to the entanglement entropy immediately \emph{before} crossing the first null singularity $\theta_1$, a geodesic with the same winding number will dominate for any $\theta>\theta_1$, compare with Figs.~\ref{fig:continuationvacuum},~\ref{fig:continuationdefect}, and the next null singularities at $\theta_i>\theta_1$ will not be detected. This shows that the saddle giving the dominant contribution to the timelike entanglement entropy after a null singularity has to be chosen by minimizing the real part of the length \emph{after} performing the analytical continuation of the relevant geodesic configurations across the light cone. This provides further support for the prescription outlined in Sec.~\ref{sec:idea}, as already anticipated at the end of Sec.~\ref{sec:black_brane_TEE}.

\section{Outlook}
\label{sec.outlook}

A significant portion of our understanding of quantum field theory phenomena occurring at temporal separations, such as subsequent measurements or a response of a system to a local perturbation, is based on analytic continuations of operator insertion points in correlation functions away from a Cauchy slice (constant time slice in some foliation). In the present paper, building on earlier developments in~\cite{Doi:2022iyj,Doi:2023zaf}, we applied the same principle to entanglement entropy in quantum field theory and defined the temporal entanglement by means of an analytic continuation of an entangling region to acquire a temporal extent. Our analytic continuation can be thought of as a generalization of kinematic space research program, see~\cite{Czech:2015qta,deBoer:2015kda,Czech:2016xec,deBoer:2016pqk}, which studies entanglement entropy dependence on the shape and location of the subregion, including how it changes as a function of both space and time. 

Within the kinematic space paradigm to date, the subregions of interest (together with their complements to be able to define a state) were bound by light cones. In the present work we propose to \emph{define} temporal entanglement entropy in Minkowski spacetimes in terms of a spacetime transformation encapsulated in Fig.~\ref{fig:rotation}. The transformation in question needs to be complexified in order to go past the light cone. This spacetime `rotation' when applied to known closed-form expressions for entanglement entropy reproduces the results of~\cite{Doi:2022iyj,Doi:2023zaf}.

However, given the scarceness of exact expressions for entanglement entropy, the key power of our idea lies in its applicability to holography, where entanglement entropy calculations amount to studying extremal surfaces in higher-dimensional spacetimes. Within our approach, all extremal surfaces that could contribute to holographic entanglement entropy are analytically continued following the change of the asymptotic boundary condition encapsulated by Fig.~\ref{fig:rotation}. The holographic timelike entanglement entropy is then computed by the resulting complex extremal surface with the smallest real part of the area.

Our holographic investigations resolve an earlier puzzle of which complex extremal surface to pick as the dominant contribution to holographic timelike entanglement entropy if multiple exist that was posed by~\cite{Heller:2024whi}: the one that minimizes the real part of the area among the complex extremal surfaces that arise as analytic continuation of candidate extremal surfaces for computing holographic entanglement entropy. Interestingly, the minimization aspect of our construction leads to strong subadditivity of holographic timelike entanglement entropy provided the real part of all involved surfaces is nonzero. It would be very interesting to understand if this subadditivity is a feature of holographic setups, or extends to the analytic continuation of entanglement entropy in general quantum field theories.

More along these lines, we also uncovered a lesson about holographic entanglement entropy itself: self-consistency of our prescription requires not considering complex extremal surfaces as possible subleading (in real part of the area) contributions to holographic entanglement entropy.

While in the present paper we studied two holographic setups that were chosen to test different aspects of our key idea, the prescription we outlined here is ready made to undertake a comprehensive exploration of temporal entanglement across the whole holographic entanglement entropy landscape. Such studies would allow to uncover detailed properties of holographic timelike entanglement entropy and, in particular, could lead to identifying phenomena for which it arises as a natural quantity to consider. The perspective that we have in mind originates from the physics of correlation functions. For example, while it is true that the shear viscosity of a quantum field theory in its thermal state is encoded in a Euclidean correlator of the energy-momentum tensor, it is much easier to access it from an analytically continued correlator: the retarded one. Although unknown to us at the moment, we expect there are phenomena when timelike entanglement entropy is in a similar vein more natural to consider than the entanglement entropy itself.

Another interesting aspect of our construction for future studies is the connection with the notion of temporal entanglement pursued in~\cite{Milekhin:2025ycm}. We believe this connection manifest itself in the two-dimensional conformal field theory setup on a Lorentzian cylinder considered in Sec.~\ref{sec.R12}. Our prescription as is does not allow to define temporal entanglement entropy for intervals of larger extent in time than the circumference of the cylinder, as this is the largest possible size of a spatial interval giving rise to a standard notion of a state in quantum field theory. However, if one were to consider spatial intervals wrapping along the cylinder and of arbitrary length, then their analytic continuation could be used as a definition of timelike entanglement entropy for timelike intervals of arbitrary extent. Upon a slight boost, such spatial intervals of arbitrary lengths are reminiscent of the approach of Ref.~\cite{Milekhin:2025ycm}, as they would be spacelike subregions containing nevertheless timelike separated points.

Finally, it would be very interesting to study the analytic continuation pursued in the present paper from the  perspective of quantum many-body systems giving rise to relativistic quantum field theories at low energies. For example, while the kinematic singularities encountered in Sec.~\ref{sec.RS1} should also be there in a regularized way on a lattice (as they are associated with the causal structure of the spacetime in which the quantum field theory lives), what we identified as the bulk singularities should not be present in a general discrete quantum many-body system. This shows that such studies have a potential of understanding which features of timelike entanglement entropy in computable examples are holography-specific and which ones might be more general and, perhaps, amenable to a general level proof or a higher level physical argument.

\begin{acknowledgments}
We would like to thank Hong Liu, Alexey Milekhin and Luca Tagliacozzo for discussions on the topic of temporal entanglement, Rob Myers for discussions and making us interested in the case studied in Sec.~\ref{sec.RS1}, and Carlos Nunez and Dibakar Roychowdhury for sharing with us their related results before they were online. This project has received funding from the European Research Council (ERC) under the European Union’s Horizon 2020 research and innovation programme (grant number: 101089093 / project acronym: High-TheQ). Views and opinions expressed are however those of the authors only and do not necessarily reflect those of the European Union or the European Research Council. Neither the European Union nor the granting authority can be held responsible for them. This work was partially supported  by the Priority Research Area Digiworld under the program Excellence Initiative  - Research University at the Jagiellonian University in Krakow. FO is supported by the Research Foundation Flanders (FWO) doctoral fellowship 1182825N. MPH would like to acknowledge hospitality of Nordita when this work was being completed.
\end{acknowledgments}

\bibliographystyle{bibstyl}
\bibliography{tlee} 

\end{document}